\documentclass[usenatbib,fleqn]{mnras}

\usepackage{mathptmx}

\usepackage{amsmath}

\usepackage{graphicx}

\usepackage{hyperref}

\usepackage{xspace}

\usepackage{color}

\newcommand{\epow}[1]{\mathrm{e}^{#1}}

\newcommand{\derivative}[2]{\frac{\mathrm{d}{#1}}{\mathrm{d}{#2}}}

\newcommand{\Mpc}{\,h^{-1}\mathrm{Mpc}}

\newcommand{\iMpc}{\,h\mathrm{Mpc}^{-1}}
\newcommand{\eV}{\,\mathrm{eV}}
\newcommand{\Msun}{\,h^{-1}\mathrm{M_\odot}}

\newcommand{\Om}{\Omega_\mathrm{m}}
\newcommand{\Ov}{\Omega_\mathrm{v}}
\newcommand{\Ob}{\Omega_\mathrm{b}}

\newcommand{\Onu}{\Omega_\nu}
\newcommand{\Oc}{\Omega_\mathrm{c}}

\newcommand{\dc}{\delta_\mathrm{c}}

\newcommand{\Dv}{\Delta_\mathrm{v}}

\newcommand{\eg}{e.g.,\xspace}
\newcommand{\ie}{i.e.\xspace}

\newcommand{\nbody}{$N$-body\xspace}
\newcommand{\LCDM}{$\Lambda$CDM\xspace}
\newcommand{\halofit}{\textsc{halofit}\xspace}
\newcommand{\mghalofit}{\textsc{mg-halofit}\xspace}
\newcommand{\pkann}{\textsc{pkann}\xspace}
\newcommand{\meadfit}{\textsc{hmcode}\xspace}
\newcommand{\emu}{\textsc{cosmic emu}\xspace}
\newcommand{\gadget}{\textsc{gadget-2}\xspace}

\newcommand{\cfhtlens}{CFHTLenS\xspace}
\newcommand{\codecs}{\textsc{CoDECS}\xspace}

\newcommand{\camb}{\textsc{camb}\xspace}
\newcommand{\mgcamb}{\textsc{mg-camb}\xspace}

\newcommand{\citecaption}[1]{\protect\cite{#1}}

\newcommand{\meadaddress}{\texttt{https://github.com/alexander-mead/hmcode}\xspace}

\newcommand{\new}[1]{{\color{black}{#1}}}

\def\mnras{MNRAS}
\def\apj{ApJ}
\def\apjl{ApJ Let.}
\def\prd{PRD}
\def\aap{A\&A}
\def\jcap{J. Cosmol. Astropart. Phys.}
\def\apjs{ApJS}

\def\m@th{\mathsurround=0pt }
\def\eqalign#1{\null\,\vcenter{\openup1\jot\m@th\ialign{\strut\hfil$\displaystyle{##}$&$\displaystyle{{}##}$\hfil\crcr#1\crcr}}\,}
\def\gtsim{\mathrel{\lower0.6ex\hbox{$\buildrel {\textstyle >}\over {\scriptstyle \sim}$}}}
\def\ltsim{\mathrel{\lower0.6ex\hbox{$\buildrel {\textstyle <}\over {\scriptstyle \sim}$}}}

\title[Extended accurate halo models]{Accurate halo-model matter power spectra with dark energy, massive neutrinos and modified gravitational forces}
\author[A. J. Mead et al.]{A. J. Mead$^{1,2,3}$\thanks{e-mail: alexander.j.mead@googlemail.com}, C. Heymans$^{3}$, L. Lombriser$^{3}$, J. A. Peacock$^{3}$, O. I. Steele$^{3}$ 
\newauthor
and H. A. Winther$^{4}$ \\
$^{1}$Department of Physics and Astronomy, University of British Columbia, 6224 Agricultural Road, Vancouver, BC, V6T 1Z1, Canada\\
$^{2}$Canadian Institute for Theoretical Astrophysics, University of Toronto, M5S 3H8, ON., Canada\\
$^{3}$Institute for Astronomy, University of Edinburgh, Royal Observatory, Blackford Hill, Edinburgh EH9 3HJ, UK\\
$^{4}$Department of Astrophysics, University of Oxford, Denys Wilkinson Building, Keble Road, Oxford, OX1 3RH, UK\\
}

\date{Accepted 2016 March 18. Received 2016 March 17; in original form 2016 February 05.}
\pagerange{\pageref{firstpage}--\pageref{lastpage}} 
\pubyear{2016}

\begin{document}
\maketitle

\label{firstpage}

\begin{abstract}
We present an accurate non-linear matter power spectrum prediction scheme for a variety of extensions to the standard cosmological paradigm, which uses the tuned halo model previously developed in \citeauthor{Mead2015b}. We consider dark energy models that are both minimally and non-minimally coupled, massive neutrinos and modified gravitational forces with chameleon and Vainshtein screening mechanisms. In all cases we compare halo-model power spectra to measurements from high-resolution simulations. We show that the tuned halo model method can predict the non-linear matter power spectrum measured from simulations of parameterized $w(a)$ dark energy models at the few per cent level for $k<10\iMpc$, and we present theoretically motivated extensions to cover non-minimally coupled scalar fields, massive neutrinos and Vainshtein screened modified gravity models that result in few per cent accurate power spectra for $k<10\iMpc$. For chameleon screened models we achieve only $10$ per cent accuracy for the same range of scales. Finally, we use our halo model to investigate degeneracies between different extensions to the standard cosmological model, finding that the impact of baryonic feedback on the non-linear matter power spectrum can be considered independently of modified gravity or massive neutrino extensions. In contrast, considering the impact of modified gravity and massive neutrinos independently results in biased estimates of power at the level of $5$ per cent at scales $k>0.5\iMpc$. An updated version of our publicly available \meadfit can be found at \meadaddress.
\end{abstract}


\section{Introduction}

%Standard
The mechanism behind the current phase of accelerated expansion of the Universe is not understood. If acceleration is caused by a cosmological constant (or vacuum energy) then the magnitude is tiny compared to expectations from fundamental physics. The alternatives are either that space is filled with dark energy (DE), an accelerant with strange properties that has never been detected in laboratories; or that the gravitational equations of \emph{Einstein} are insufficient and that acceleration arises naturally within the correct theory, so-called modified gravity (MG) theories. There are now many models that purport to explain accelerated expansion, and often these models make different predictions for how matter is clustered in the Universe. This clustering can be probed via `weak' lensing: measuring a small, correlated distortion of galaxy shapes that is imparted by the light-bending effect of matter along photon trajectories between the galaxy and a telescope on Earth. It follows that the lensing signal can be used to infer the matter clustering and thus discriminate between models. %Therefore, weak lensing can distinguish between models by measuring clustering, irrespective of luminosity, and from this one can infer properties of the accelerated expansion of the cosmos. The fact that the weak lensing signal is sensitive to all matter makes it attractive for constraining cosmological parameters when compared to measurements of biased tracers of the matter, such as galaxies.

%Weak lensing disadvantages
Unfortunately, weak lensing has its disadvantages: foremost among these is that it measures a projected version of the density field, where difficult-to-model, non-linear, small-scale perturbations close to the observer contribute to the shear signal at the same angular scale as larger scale perturbations further away. This angular scale mixing means that the clean, linear signal is difficult to isolate (although isolation can be partially achieved using tomographic or full 3D lensing \citealt{Heavens2003,Heymans2013,Kitching2014}). However, there is significant information in non-linear scales that are automatically measured by lensing, and this information can be exploited with accurate enough modelling. In principle, this accuracy \emph{is} attainable using cosmological \nbody simulations to determine the small-scale clustering, but simulations are too slow to be run for every combination of model parameters under consideration, and this problem is only amplified when one considers the vast number of different models that may be constrained via lensing.

%Dark energy
DE models provide an alternative to the standard cosmological constant to bring about the accelerated expansion of the cosmos. Typically, this is achieved by adding a new component with exotic properties to the matter--energy budget of the Universe. These models are dynamical, with a field whose value changes with time. In minimally coupled models DE only clusters at the horizon scale, but structure formation on smaller scales is indirectly affected by the modified background expansion rate. In non-minimally coupled models, there can be direct interactions between DE and (dark) matter. Comprehensive reviews of DE models are given by \cite{Copeland2006} and \cite{Amendola2010}.

%Neutrinos
The neutrino mass may conceivably be constrained by weak lensing, due to the effect of massive neutrinos on structure formation. Flavour oscillations \citep{Fukuda1998,SNO2002} point definitively to non-zero masses for at least two of the three neutrino mass eigenstates, but oscillation experiments measure mass--square differences, rather than absolute masses. This puts a lower bound on the sum of neutrino masses of $\simeq0.06\eV$ (normal hierarchy) or $\simeq0.1\eV$ (inverted hierarchy). Forthcoming cosmological surveys will be powerful enough to distinguish between these two possibilities \cite[\eg][]{Hall2012}. Neutrinos with masses that are allowed by current cosmological data are relativistic when they decouple, and at recombination, but become non-relativistic later ($z_\mathrm{nr}\simeq 2000\,m_\nu\,\eV^{-1}$). They initially behave like radiation, with no sub-horizon clustering ($\bar\rho_\nu \propto a^{-4}$), but later transition to behaving like matter ($\bar\rho_\nu \propto a^{-3}$) and are able to cluster. Thus, neutrinos alter both the background expansion of the Universe (and the angular-diameter distance relation) and also the growth of cold-matter perturbations directly, because neutrinos are smoothly distributed below their free-streaming scale and so do not contribute to structure development small scales. Reviews of neutrinos in cosmology are given by \cite{Lesgourgues2006} and \cite{Abazajian2015}.

%Paragraph on MG
A MG force law is one way to explain the accelerated expansion of the Universe. In contrast to DE models, MG models attempt to give rise to accelerated expansion via a direct modification of the gravitational force law. In order to comply with accurate tests of gravity on Earth, or at the scale of the Solar system, these models require a `screening' mechanism, so that gravity is restored to the standard in environments where it is well tested, while still being allowed to behave differently on large scales and perhaps to give rise to accelerated expansion. In chameleon theories \citep[\eg][]{Khoury2004} the screening is primarily a function of halo mass and environment, while in \cite{Vainshtein1972} models the screening depends primarily on the local density. Recent reviews of some of the menagerie of possible MG theories are given by \cite{Joyce2015} and \cite{Koyama2016}.

%The point of this paper
In this paper, we provide a halo model based fitting function for the power spectrum of matter fluctuations for a subset of extensions to the standard cosmological paradigm. We build on the work presented in \cite{Mead2015b} in which we developed a version of the standard halo model \citep{Peacock2000,Seljak2000,Cooray2002} that produces matter power spectra that are accurate at the $\simeq 5$ per cent level for a range of standard cosmological parameters ($\Om$, $\Ob$, $\sigma_8$, $w$, $h$, $n_\mathrm{s}$). This accuracy was achieved by treating several parameters in the usual halo model as free, and fitting these to data from the high resolution simulations of the \emu collaboration \citep{Heitmann2009,Heitmann2010,Lawrence2010,Heitmann2014}. \emu was chosen because the simulations were high precision and thoroughly tested for resolution issues, as well as being designed to cover a large cosmological parameter space. In \cite{Mead2015b} we also investigated the effects of baryonic feedback on the matter spectrum, and showed that the calibrated halo model could be extended to provide a single parameter fitting recipe for a range of baryonic feedback models. Our halo model has recently been used by \cite{Joudaki2016} to analyse data from the \cfhtlens survey to provide cosmological constraints while marginalizing over possible feedback models. In this paper, we focus on DE models, massive neutrinos and MG force laws, which are not part of the \emu parameter space. Our results are therefore presented at the level of the power spectrum response (ratio of power in the new model to some fiducial model, which we consider to be well described by our previous work) predicted by the halo model, compared to the same response seen in simulations. The fitting formula we produce in this paper can be used in forthcoming weak lensing data analyses to constrain the additional model parameters, or for any application that requires a matter power spectrum.

%This paper is arranged...
This paper is arranged as follows. In Section~\ref{sec:halomodel}, we briefly summarize the basics of the halo model from \cite{Mead2015b} that we use to calculate accurate power spectra. We then present our results in Section~\ref{sec:results}: for the power spectrum of DE (Sections \ref{sec:de}, \ref{sec:codecs}), massive neutrinos (Section \ref{sec:massive-nu}) and MG models (Sections \ref{sec:chameleon}, \ref{sec:Vainshtein}). In Section \ref{sec:degeneracy}, we briefly investigate degeneracies between some of the extensions to the standard cosmological paradigm that we have considered, and comment on the viability of treating these extensions separately. Finally, we summarise in Section~\ref{sec:summary}.

\section{Halo model power spectra}
\label{sec:halomodel}

Weak lensing measures a projected version of the 3D density distribution. Therefore we are primarily interested in the power spectrum of the 3D matter over-density $\delta$, defined relative to the background density via $1+\delta=\rho/\bar\rho$. The power spectrum of statistically isotropic density fluctuations depends only on $k=|\mathbf{k}|$ and is given by
\begin{equation}
P(k)=\langle|\delta_\mathbf{k}|^2\rangle\ ,
\label{eq:P(k)}
\end{equation}
where the average is taken over modes with the same modulus but different orientations. We find it more convenient to use the dimensionless quantity $\Delta^2$:
\begin{equation}
\Delta^2(k)=4\pi L^3\left(\frac{k}{2\pi}\right)^3 P(k)\ ,
\label{eq:Delta2_definition}
\end{equation}
which gives the fractional contribution to the variance per logarithmic interval in $k$ in a cube of volume $L^3$. If the over-density field is filtered on a comoving scale $R$, the variance is
\begin{equation}
\sigma^2(R,z)=\int_0^{\infty}\Delta^2(k,z)\, T^2(kR)\;\mathrm{d}\ln{k}\ .
\label{eq:variance}
\end{equation}
If instead the 1D linear displacement field is filtered then the variance is
\begin{equation}
\sigma_\mathrm{d}^2(R,z)=\frac{1}{3}\int_0^{\infty}\frac{\Delta^2(k,z)}{k^2}\, T^2(kR)\;\mathrm{d}\ln{k}\ .
\label{eq:sigma_v}
\end{equation}
In both cases we use the window function
\begin{equation}
T(x)=\frac{3}{x^3}(\sin{x}-x\cos{x})\ ,
\label{eq:top_hat}
\end{equation} 
corresponding to smoothing with a spherical top-hat. Note that $\sigma(R)$ is dimensionless, whereas $\sigma_\mathrm{d}(R)$ has dimensions of length, $\sigma(R\to\infty)$ diverges for standard cosmological spectra, but $\sigma_\mathrm{d}(R\to 0)$ is well defined and we denote this asymptotic value as $\sigma_\mathrm{d}$ with no argument.

In this paper, we use the model previously developed in \cite{Mead2015b} and associated code\footnote{\url{https://github.com/alexander-mead/hmcode}} \citep{HMcode}: a variant of the standard halo model with parameters tuned to match the power spectra data from the \emu simulations. In that paper, we showed that it was possible to fit a range of cosmological models at the $5$ per cent level for $z\leq1$  and 10 per cent level for $1<z\leq2$ for $k\leq10\iMpc$. We refer the reader to that paper for full details of our halo model, but we provide a short summary here.

\begin{table*}
\caption{Halo-model parameter descriptions and values before and after fitting. We show standard values, those from \citeauthor{Mead2015b} (\citeyear{Mead2015b}), and updated values. A dash in `updated value' column indicates no change from the \citeauthor{Mead2015b} (\citeyear{Mead2015b}) value. $\sigma_{d,100}(z)=\sigma_\mathrm{d}(R=100\Mpc,z)$ with $\sigma_\mathrm{d}$ defined in equation~(\ref{eq:sigma_v}).}
\begin{center}
\begin{tabular}{c c c c c c}
\hline
Parameter & Description & Standard value & \cite{Mead2015b} value & Updated value & Equations in text \\
\hline
$\Dv$ & Virialized halo overdensity & 200 & $418\,\Om^{-0.352}(z)$ & -- & \ref{eq:Dv_definition} \\
$\dc$ & Linear collapse threshold & 1.686 & $1.59+0.0314\,\ln\sigma_8(z)$ & -- & \ref{eq:dc_definition}, \ref{eq:dc_nakamura} \\
$\eta$ & Halo bloating parameter & 0 & $0.603-0.3\,\sigma_8(z)$ & -- &\ref{eq:1haloterm} \\
$f$ & Linear spectrum damping factor & 0 & $0.188\,\sigma_8^{4.29}(z)$ & $0.0095\,\sigma^{1.37}_{d,100}(z)$ & \ref{eq:2haloterm} \\
$k_*$ & One-halo damping wavenumber & 0 & $0.584\,\sigma_\mathrm{d}^{-1}(z)$ & -- & \ref{eq:1haloterm} \\
$A$ & Minimum halo concentration & 4 & $3.13$ & -- & \ref{eq:bullock_cm} \\
$\alpha$ & Quasi-linear one- to two-halo softening & 1 & $2.93\times 1.77^{n_\mathrm{eff}}$ & $3.24\times1.85^{n_\mathrm{eff}}$ & \ref{eq:meadfit} \\
\hline
\end{tabular}
\label{tab:fit_params}
\end{center}
\end{table*}

In the simple case of Poisson distributed spherical haloes the power spectrum has the form of shot noise, moderated by the density profile of the haloes. We augment this to provide accurate spectra in the following way:
\begin{equation}
\eqalign{
\Delta_\mathrm{1H}^2(k)=&[1-\epow{-(k/k_*)^2}]4\pi\left(\frac{k}{2\pi}\right)^3\frac{1}{\bar\rho}\cr&\times\int_0^\infty M(\nu) W^2(\nu^\eta k,M) f(\nu)\;\mathrm{d}\nu\ .
}
\label{eq:1haloterm}
\end{equation}
The one-halo term is calculated as an integral over all halo masses, $M$, whose mass relates to the peak threshold $\nu$ via
\begin{equation}
\nu=\dc(z)/\sigma(M,z)\ ,
\label{eq:dc_definition}
\end{equation}
where $f(\nu)$ is the halo mass function, which can be expressed as a near-universal function \citep[\eg][]{Press1974,Sheth1999,Jenkins2001} in $\nu$. Usually $\dc$ is fixed at a constant $\simeq1.686$, in accordance with $\Om=1$ spherical-collapse model predictions, but in \cite{Mead2015b} we fitted $\dc$ and endowed it with some redshift dependence. Since publishing \cite{Mead2015b}, we found that slightly improved results for halo-model power spectra were obtained if we used the predicted variation of $\dc$ from the spherical collapse model for $\Lambda$ cold dark matter (\LCDM). Therefore our $\dc$ is now augmented by the fitting formula of \cite{Nakamura1997}:
\begin{equation}
\dc= [1.59+0.0314\ln\sigma_8(z)]\times[1+0.0123\log_{10}\Om(z)]\ .
\label{eq:dc_nakamura}
\end{equation}
Even though this change to $\dc$ is very small, the effects are felt because $\dc$ is exponentiated in the mass function. In equation~(\ref{eq:1haloterm}), $W(k,M)$ is the normalized Fourier transform of the halo-density profile:
\begin{equation}
W(k,M)=\frac{1}{M}\int_0^{r_\mathrm{v}}\frac{\sin(kr)}{kr}\ 4\pi r^2\rho(r,M)\;\mathrm{d}r\ , 
\label{eq:halo_window}
\end{equation}
where $r_\mathrm{v}$ is the halo virial radius. We damp the one-halo term at small $k$ to prevent it from becoming larger than linear on very large scales (which is unphysical), which is governed by the fitted parameter $k_*=0.584\sigma_\mathrm{d}^{-1}(z)$. In addition, we use the fitted parameter $\eta=0.603-0.3\sigma_8(z)$ in equation~(\ref{eq:1haloterm}) to bloat or constrict haloes as a function of their mass at fixed virial radius. Haloes are considered to be objects that are $\Dv$ times denser than the background, which implies
\begin{equation}
M=\frac{4}{3}\pi r_\mathrm{v}^3 \Dv\bar\rho\ .
\label{eq:Dv_definition}
\end{equation}
In \cite{Mead2015b}, we fitted for $\Dv$ and obtained $\Dv=418\,\Om^{-0.352}(z)$. 

For the mass function we use the formula of \cite{Sheth1999}, which is an empirical fit to simulations:
\begin{equation}
f(\nu)=A\left[1+\frac{1}{(a\nu^{2})^p}\right]\epow{-a\nu^2/2}\ ,
\label{eq:STmassfunction}
\end{equation}
where the parameters of the model are $a=0.707$, $p=0.3$ and $A$ is constrained by the property that the integral of $f(\nu)$ over all $\nu$ must equal one: $A\simeq 0.2162$. We use halo profiles of  \citeauthor*{Navarro1997} (NFW; \citeyear{Navarro1997}):
\begin{equation}
\rho(r)=\frac{\rho_\mathrm{N}}{(r/r_\mathrm{s})(1+r/r_\mathrm{s})^2}\ ,
\label{eq:nfw}
\end{equation}
where $r_\mathrm{s}$ is a scale radius that roughly separates the core of the halo from the outer portion and $\rho_\mathrm{N}$ is a normalization. The scale radius is typically expressed via the halo concentration, $c=r_\mathrm{v}/r_\mathrm{s}$. We use the concentration--mass relations of \cite{Bullock2001}:
\begin{equation}
c(M,z)=A\frac{1+z_\mathrm{f}(M)}{1+z}\left[\frac{g(z\to\infty)}{g_{\Lambda}(z\to\infty)}\right]^{1.5}\ , 
\label{eq:bullock_cm}
\end{equation} 
where the ratio of linear growth functions\footnote{$g(z)$ is normalized such that $g(z=0)=1$} is a correction advised by \cite{Dolag2004} for DE models. In \cite{Mead2015b}, we did not have the $1.5$ exponent in this correction, but we subsequently discovered that this produced more accurate power spectra for the more extreme DE models. The calculation of the halo formation redshift, $z_\mathrm{f}$, as a function of mass, is described in \cite{Bullock2001}, and crucially depends on the formation history of the halo, so there is some hysteresis whereby haloes retain a memory of their formation time. 

On large scales haloes are not Poisson distributed, and displacements of haloes with respect to one another require us to add a `two-halo' term to the power.  For the matter distribution this is approximately the linear-theory power spectrum, however, perturbation theory \cite[\eg][]{Crocce2006a} suggests this should be damped at quasi-linear scales, and so we work with a perturbation theory inspired two-halo term with damping that asymptotes to a constant at small scales
\begin{equation}
\Delta^{'2}_\mathrm{2H}(k)=\left[1-f\tanh^2{(k\sigma_\mathrm{d}/\sqrt{f})}\right]\Delta^2_\mathrm{lin}(k)\ ,
\label{eq:2haloterm}
\end{equation}
where $f$ is a fitted parameter (see Table~\ref{tab:fit_params}).

Usually the expression for the full halo-model power spectrum is given by a simple sum of the one- and two-halo terms, but in \cite{Mead2015b} we found that it was necessary to smooth this transition to account for the well known deficit in power in the transition region. We instead use
\begin{equation}
\Delta^2(k)=[(\Delta_\mathrm{2H}^{'2})^\alpha+(\Delta_\mathrm{1H}^{'2})^\alpha]^{1/\alpha}\ ,
\label{eq:meadfit}
\end{equation}
where $\alpha$ is a fitted parameter (see Table~\ref{tab:fit_params}), which depends on the effective spectral index at the non-linear scale, defined as
\begin{equation}
3+n_\mathrm{eff}(z)\equiv \left.-\frac{\mathrm{d}\ln\sigma^2(R,z)}{\mathrm{d}\ln R}\right\vert_{\sigma=1}\ .
\label{eq:n_eff_definition}
\end{equation}
In this work, we consider extensions to the $w$CDM paradigm, and to accommodate these we find it necessary to modify some of our fitted parameters. This refitting actually slightly improves the results presented in \cite{Mead2015b}. Our updated values given in Table \ref{tab:fit_params} and described in the text at the appropriate points.

\section{Results}
\label{sec:results}

%How the results section is organised
In this section, we present our results for different beyond standard-model paradigms. In each case, we compare simulated power spectra to spectra predicted by our augmented halo model at the level of the ratio of each prediction to a reference case, which is always a vanilla \LCDM model. This `response' does not measure the absolute accuracy of any method of power spectrum prediction, but instead demonstrates how well a method responds to the changes of the new paradigm. Computing the response from simulations that share an initial random seed has the additional advantage of cancelling cosmic variance and some resolution artefacts \citep[\eg][]{McDonald2006,Smith2014}. In \cite{Mead2015b}, we demonstrated the absolute accuracy of our augmented halo model and we direct the reader to that paper for more information. 

%Halofit
For each new paradigm, we also compare the power spectrum response as predicted by various incarnations of \halofit: a fitting function that is commonly used to produce the non-linear power spectrum that was originally developed by \cite{Smith2003}. \halofit was trained to accurately reproduce non-linear power for a wide range of cosmological models, including those with power-law initial spectra as well as Einstein--de--Sitter, \LCDM and open models with standard curved initial spectra. \halofit was updated by \cite{Takahashi2012} where some parameters were re-tuned so that the fit was more accurate at smaller scales around the smaller range of cosmological parameter space preferred by contemporary observational data. In addition, the re-tuning involved simulations with constant $w=p/\rho$ DE, and this $w$ appears explicitly in the revised fitting formula. \halofit can be used to produce non-linear spectra for any model, all that is required is an input linear spectrum together with values of the cosmological density parameters. However, there is no guarantee of accuracy for models that are `far' from where \halofit was trained. For massive neutrino models, \cite*{Bird2012} provided an update with new \halofit parameters that depend explicitly on the neutrino mass, and \cite{Zhao2014} provided a similar update for $f(R)$ chameleon gravity models.

\subsection{DE: quintessence}
\label{sec:de}

%\cite{Alimi2010} % Imprints of DE I - imprints on P(k)
%\cite{Courtin2011} % Imprints of DE II - non-universality of hmf - dc(z), Dv(z) does help
%\cite{Jennings2010} %QUICC simulations
%\cite{McDonald2006} %DE sims with fixed linear theory

\begin{table}
\begin{center}
\caption{Parameters of the $w(a)$CDM simulations. All simulations use $512^3$ particles in cubes of size $L=200\Mpc$ and start from identical initial conditions. The initial transfer function was generated using \camb with cosmological parameters $\Om=0.307$, $\Ov=1-\Om$, $\Ob=0.0483$, $h=0.678$, $n_\mathrm{s}=0.961$.}
\begin{tabular}{c c c c}
\hline
Model & $w_0$ & $w_a$ & $\sigma_8$ \\
\hline
\LCDM & $-1$ & $0$ & $0.844$ \\
VLOW & $-1$ & $-1$ & $0.888$ \\
LOW & $-1$ & $-0.5$ & $0.871$ \\
HIGH & $-1$ & $0.5$ & $0.794$ \\
VHIGH & $-1$ & $0.7$ & $0.754$ \\
\hline
\end{tabular}
\label{tab:de}
\end{center}
\end{table}

DE models provide an alternative to the standard cosmological constant to bring about the accelerated expansion of the Cosmos and are dynamical, with a field whose value changes with time. The simplest possible models involve a single scalar field with a self-interaction potential (often called quintessence; \new{\citealt{Ratra1988,Wetterich1988b}}) and that are minimally coupled to gravity via $\sqrt{|g|}$ that appears in the action. Additional complexity is possible by adding multiple fields, a non-standard kinetic term, or by considering fields other than scalars. The motivation behind many of these models is that they may explain the `why now' aspect of the cosmological constant problem because certain scalar field models have been shown to demonstrate `tracking' behaviour whereby the scalar energy-density tracks that of the dominant component in the background \citep*{Copeland1998,Barreiro2000}. It may be that such a scalar field started life tracking the radiation density, and that the radiation-to-matter transition changed the field to a state where it causes accelerated expansion in the late Universe.

For DE models, it is usual to express the pressure in terms of the energy density via $p=w\rho$. A cosmological constant has $w=-1$ and simple scalar-field DE models have time-varying $w$ with the constraint $-1\leq w\leq 1$. One can consider models with constant $w$, but constant $w\neq -1$ is strongly constrained by observations: \cite{PlanckXIII2015} combine cosmic microwave background (CMB) and SN1a to give constraints of $w=-1.006\pm 0.045$ ($2\sigma$). In \cite{Mead2015b} we fitted our halo model to the simulations of \emu that included models with constant $-1.3<w<-0.7$. Here, we investigate models with a time-varying equation of state of the form
\begin{equation}
w(a)=w_0+(1-a)w_a\ ,
\label{eq:wa_definition}
\end{equation}
which has been demonstrated to capture much of the phenomenology of a wide range of quintessence models \citep{Chevallier2001,Linder2003}. If $w_a=0$, this reduces to the constant $w$CDM case. \cite{Linder2015} showed that a broad range of thawing models can be accommodated with a fixed relation, $w_a\simeq 1.58(1+w_0)$. In general, dynamical models are less constrained than constant $w$ models, with acceptable ranges given by \citeauthor{PlanckXIV2015} (\citeyear{PlanckXIV2015}; fig. 4); approximately $-1.2\ltsim w_0\ltsim-0.7$ and $-1.5\ltsim w_a\ltsim0.5$.

To test the accuracy of our halo model we created five cosmological \nbody simulations, all with $w_0=-1$ and $w_a=-1$, $-0.5$, $0$, $0.5$ and $0.7$ that cover some of the $w_a$ range allowed by \cite{PlanckXIV2015}. Each simulation was run in a cube with $L=200\Mpc$, particle number $N=512^3$, and with softening set at $1/50$ the mean inter-particle separation\footnote{The other non-physical simulation parameters are: \textsc{pmgrid}=1024; $z_i=199$; grid initial conditions; $\textsc{MaxSizeTimestep}=0.01$; $\textsc{ErrTolIntAccuracy}=0.025$; $\textsc{ErrTolTheta}=0.7$; $\textsc{ErrTolForceAcc}=0.005$}. Our simulations were run using a modified version of the \gadget code \citep{Springel2005b} that can accept DE of the form given in equation (\ref{eq:wa_definition}). We verified our modifications at the level of the power spectrum by confirming that the correct linear growth of perturbations was seen at the largest scales in our simulations. Initial transfer functions were generated using \camb \citep{Lewis2000} and all simulations start from identical initial conditions. In the $w_a=0$ \LCDM model $\sigma_8=0.844$, but in the other models, perturbations have grown by different amounts due to the time-varying nature of the DE: parameters are given in Table~\ref{tab:de}.

\begin{figure}
\begin{center}
\includegraphics[angle=270,width=8.4cm]{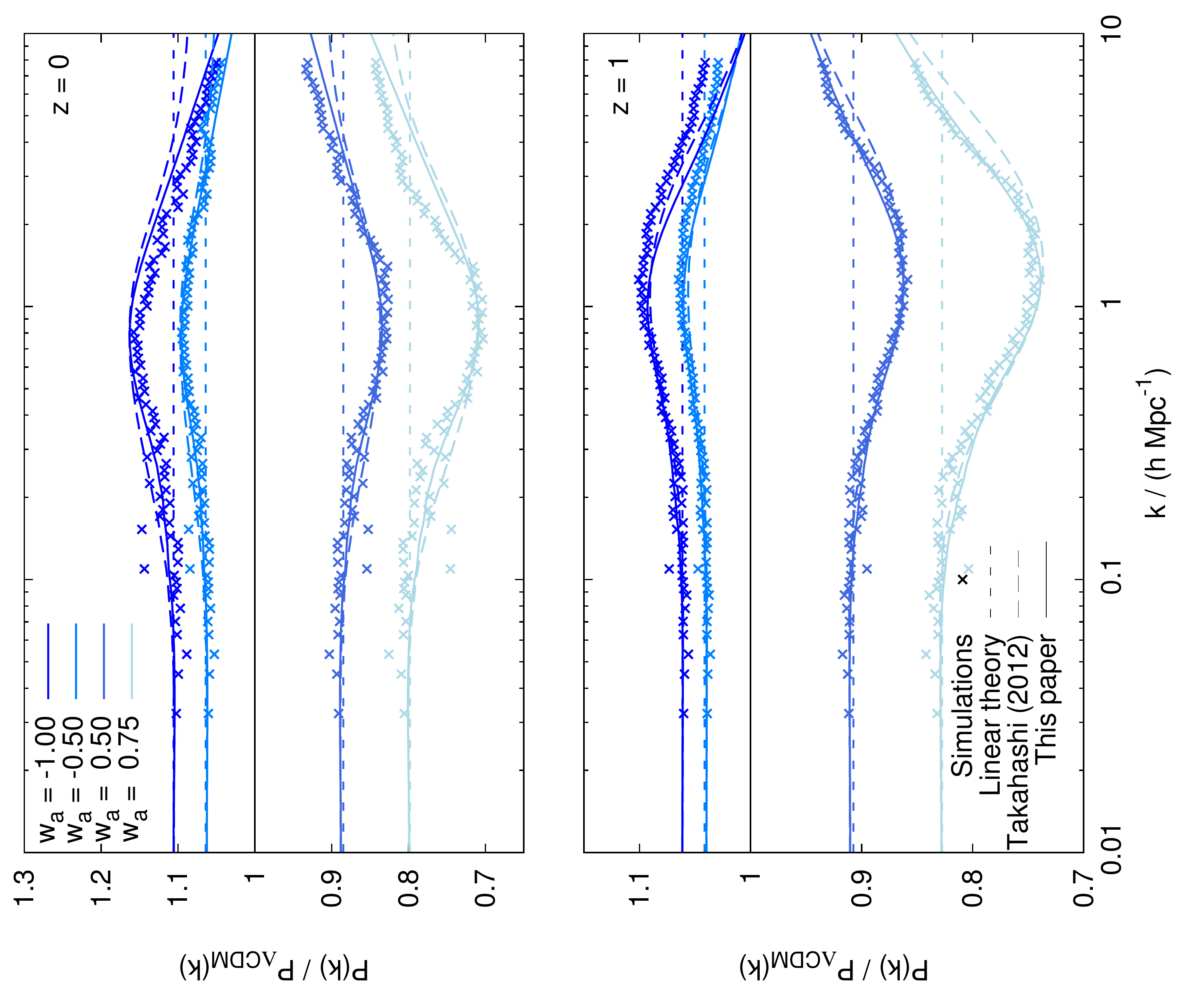}
\end{center}
\caption{The power spectra of four different DE cosmologies, compared to a \LCDM model at $z=0$ (top) and $1$ (bottom). We show power from models with $w_a=-1$ (highest curve; darkest blue), $-0.5$, $0.5$ and $0.7$ (lowest curve; lightest blue) as measured in simulations (filled circles) and as predicted by linear theory (short-dashed; flat lines), the \citecaption{Mead2015b} halo model (solid) and the \citecaption{Takahashi2012} \halofit (long-dashed). Both non-linear models predict the trend seen in simulations well.}
\label{fig:de}
\end{figure}

In Fig. \ref{fig:de}, we show the power spectrum response as measured in our $w(a)$ simulations and as predicted by the \cite{Mead2015b} halo model and the \halofit model of \cite{Takahashi2012} at $z=0$ and $1$. We see that both models do an excellent job (few per cent accuracy) at predicting the response in non-linear power to the dynamical DE. This is probably because the DE is homogeneous in the simulations\footnote{In reality scalar field DE does cluster, but only on scales comparable to the horizon, far larger than typical simulation boxes.} and only affects the background evolution; many quantities of relevance to non-linear clustering (such as the mass function) have been shown to be expressible as near-universal functions of $\sigma(R)$ \citep{Sheth1999}, and this is incorporated into both models. This explains the accuracy to which both methods predict the `bump' (anti-bump) that develops around $k=0.2\iMpc$, which arises due to the boosted (depleted) halo population in models with enhanced (decreased) linear amplitude. The DE parameter $w$ appears explicitly in \halofit of \cite{Takahashi2012} because it was fitted to $w$CDM models. Therefore, we have a choice to set this to $w_0$ or $w(a)$ in dynamical models; we found that more accurate \halofit results were obtained in the latter case and these are the results we present. The halo model knows about the evolution of $w(a)$ through linear growth as well as the $c(M)$ relation, which is calculated using the prescription of \cite{Bullock2001} and depends on the linear growth history. This may explain the improved performance over \halofit around $k=5\iMpc$. In view of this result, we merely present these results and do not attempt to optimize any of our halo-model parameters further.

The non-linear power spectrum has been investigated via simulations in more exotic quintessence models \citep[\eg][]{Alimi2010,Jennings2010} and in the future it would be interesting to compare our halo model to simulations, although Fig.~\ref{fig:de} suggests that results should be excellent. Although \emu only investigate models with constant $w$, recently \cite{Casarini2016} have demonstrated that the \emu models may be remapped to time-varying DE models by equating a $w(a)$ model with a cosmology with identical spectral shape and amplitude as well as conformal time, which is nearly equivalent to matching the growth history, and achieved by fixing $w$. Spherical collapse models in DE have been investigated \citep[\eg][]{Percival2005}, and compared to simulations. For example, \cite{Courtin2011} show that including the predicted DE dependence of $\dc$ provides better matches to the mass function, but only at the high-mass end, and we discuss this more in Section~\ref{sec:summary}. We do not attempt to include any spherical model motivated corrections at this stage.

\subsection{\textsc{CoDECS}: non-minimally coupled scalar fields}
\label{sec:codecs}

\begin{figure*}
\begin{center}
\includegraphics[angle=270,width=17.5cm]{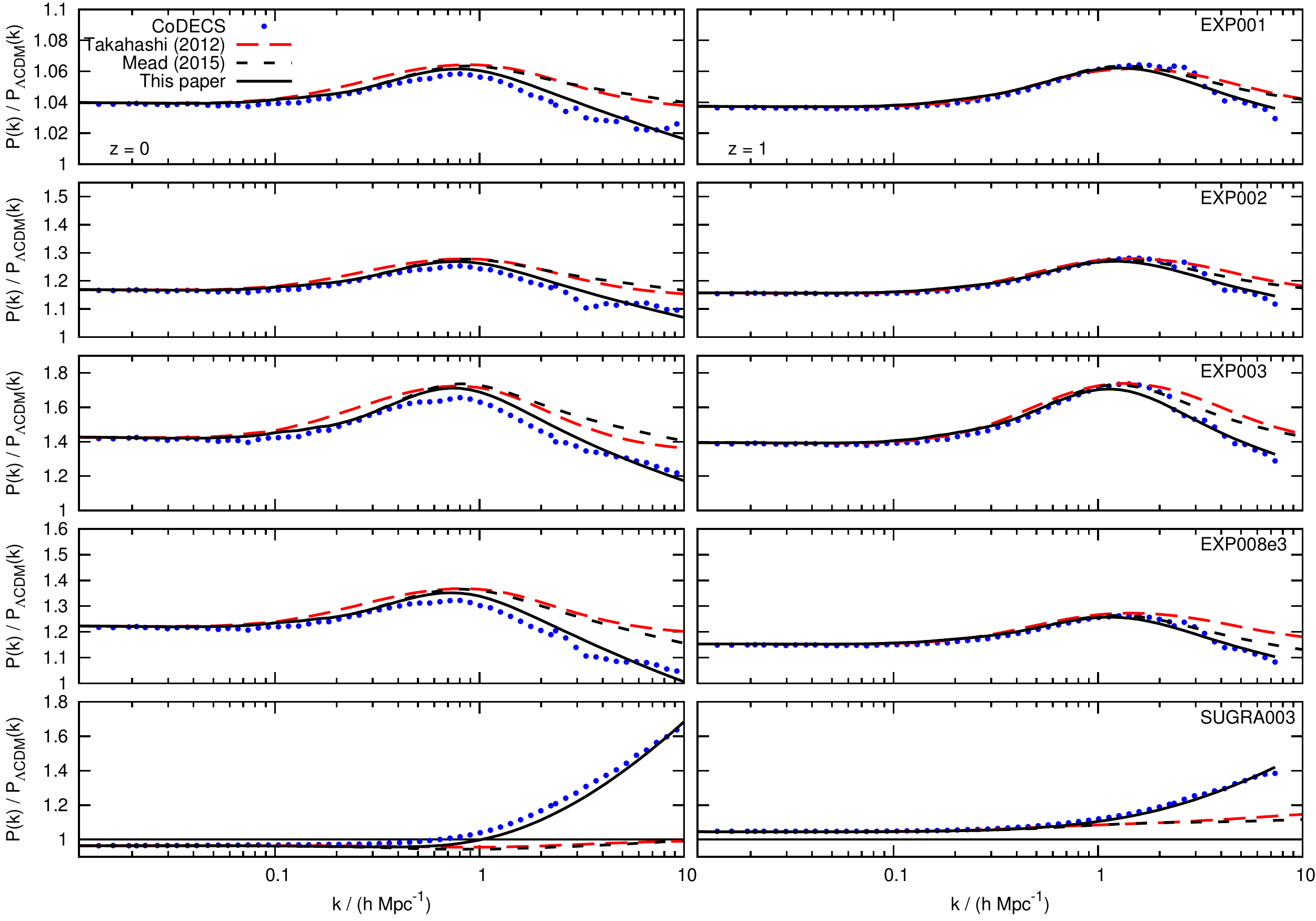}
\end{center}
\caption{The power spectrum response of DE models explored by the \codecs simulation project of \citecaption{Baldi2011}. The response measured in simulations spectra (blue points) is shown compared to halo model of \citeauthor{Mead2015b} (\citeyear{Mead2015b}; short-dashed; black), \halofit (long-dashed; red) and our updated model (solid; black) at $z=0$ (left) and $1$ (right). In the update, we have fitted the concentration--mass relation in the halo model to the simulation data, at each $z$, but we have not predicted these a priori.}
\label{fig:codecs}
\end{figure*}

%Introduction
The next paradigm we consider is that of non-minimally coupled scalar fields \citep{Wetterich1995}. These models are an attempt to unify the dark sector and attempt to mitigate the cosmological constant problem by introducing a coupling between dark matter (DM) and DE; thus it is possible that DE arises through the transfer of energy from DM to DE. %We consider models that may be derived from the 

These models were first simulated by \cite{Baldi2010} and we use publicly available data from the \codecs simulation project\footnote{\url{http://www.marcobaldi.it/web/CoDECS.html}} \citep{Baldi2012b} for scalar field DE models with a CDM--DE coupling, so that energy can flow between these two components of the cosmos. The continuity equations in the background are:
\begin{equation}
\eqalign{
\ddot\phi+3H\dot\phi+\derivative{V}{\phi}&=\sqrt{\frac{16\pi G}{3}}\beta(\phi)\rho_\mathrm{c}\ ,\cr
\dot\rho_\mathrm{c}+3H\rho_\mathrm{c}&=-\sqrt{\frac{16\pi G}{3}}\beta(\phi)\rho_\mathrm{c}\dot\phi\ ,\cr
\dot\rho_\mathrm{b}+3H\rho_\mathrm{b}&= 0\ ,
}
\end{equation}
where $\phi$ is the scalar field, $V$ is the scalar-field potential, $\rho_\mathrm{c,b}$ are the background densities of CDM and baryons, and $\beta(\phi)$ is a coupling function. Note that $\rho_\mathrm{b}\propto a^{-3}$ as usual, but this is \emph{not} the case for the CDM because of the energy flow to and from the scalar field. Baryons are unaffected by the flow of energy, which means these models are able to pass conventional gravity tests. If $\beta\dot\phi>0$ energy is transferred from CDM to DE and vice-versa.

\begin{table}
\begin{center}
\caption{Parameters of the \codecs coupled DE models. $A$ and $\alpha$ relate to the potential defined in equations~(\ref{eq:codecs_potential}; EXP models) and~(\ref{eq:codecs_sugra}; SUGRA model), while $\beta_0$ and $\beta_1$ relate to the coupling function defined in equation~(\ref{eq:codecs_coupling}). With the exception of the SUGRA model all have $w\simeq-1$ at $z=0$, but due to the different growth histories have different $\sigma_8$ at $z=0$, despite all models sharing initial conditions. For the SUGRA003 model, we note that there is a small difference in the $\sigma_8$ quoted in the \codecs papers (0.806) compared to that we infer from the ratio of power spectra (0.794) at large scales. We use the $\sigma_8$ value inferred from this ratio in our \halofit and halo model calculations for our results, which ensures that the large-scale response in Fig.~\ref{fig:codecs} is accurate.}
\begin{tabular}{c c c c c c c}
\hline
Model & A & $\alpha$ & $\beta_0$ & $\beta_1$ & $w(z=0)$ & $\sigma_8$ \\
\hline
\LCDM & $0.0219$ & -- & -- & -- & $-1$ & $0.809$ \\
EXP001 & $0.0218$ & $0.08$ & $0.05$ & $0$ & $-0.997$ & $0.825$ \\
EXP002 & $0.0218$ & $0.08$ & $0.1$ & $0$ & $-0.995$ & $0.875$ \\
EXP003 & $0.0218$ & $0.08$ & $0.15$ & $0$ & $-0.992$ & $0.967$ \\
EXP008e3 & $0.0217$ & $0.08$ & $0.4$ & $3$ & $-0.982$ & $0.895$ \\
SUGRA003 & $0.0202$ & $2.15$ & $-0.15$ & $0$ & $-0.901$ & $0.806$ \\
\hline
\label{tab:codecs}
\end{tabular}
\end{center}
\end{table}

%Explanation of model parameters
The model has free functions in $V(\phi)$ and $\beta(\phi)$. The \codecs project considers two choices for the potential, an exponential \citep[\eg][]{Wetterich1988a}
\begin{equation}
V(\phi)=A\epow{-\alpha\phi}\ ,
\label{eq:codecs_potential}
\end{equation}
and a super-gravity potential \citep{Brax1999}
\begin{equation}
V(\phi)=A\phi^{-\alpha}\epow{\phi^2/2}\ .
\label{eq:codecs_sugra}
\end{equation}
The coupling is taken to have the exponential form
\begin{equation}
\beta(\phi)=\beta_0\epow{\beta_1\phi}\ .
\label{eq:codecs_coupling}
\end{equation}
We refer the reader to \cite{Amendola2000} and \cite{Baldi2011,Baldi2012a} for a detailed discussion of these models. The names and parameters of the \codecs models we investigate are given in Table~\ref{tab:codecs}.

%Linear perturbation theory
Linear perturbations in baryons evolve according to the standard equation, but those in CDM are affected by the energy exchange. On sub-horizon scales, the perturbation equations are:
\begin{equation}
\eqalign{
\ddot\delta_\mathrm{b}+2H\dot\delta_\mathrm{b}&=4\pi G(\rho_\mathrm{b}\delta_\mathrm{b}+\rho_\mathrm{c}\delta_\mathrm{c})\ ,\cr
\ddot\delta_\mathrm{c}+2H\left[1-\frac{\beta\dot\phi}{\sqrt{6}H}\right]\dot\delta_\mathrm{c}&=4\pi G\left[\rho_\mathrm{b}\delta_\mathrm{b}+\rho_\mathrm{c}\delta_\mathrm{c}\left(1+\frac{4}{3}\beta^2\right)\right]\ .
}
\end{equation}
The friction term involving $\beta$ on the left-hand side of the CDM equation arises as a consequence of momentum conservation, whereas the $\beta^2$ term on the right-hand side is a direct force term from the scalar field. In addition, linear perturbations evolve differently because of the altered background expansion. This is in contrast to a minimally coupled quintessence model ($\beta=0$), where small-scale differences arise \emph{only} because of the modified background expansion. Note that DE should also cluster in this model, but this is ignored in both our linear perturbation theory and the simulations because this clustering is of order the gravitational potential, $\mathcal{O}(10^{-6})$, and is therefore negligible.
%Does phi move at c?

%Lucas thinks I should only mention this once.
%In the case of the SUGRA003 model we noticed a small discrepancy between the $\sigma_8$ quoted by \cite{Baldi2012b} and that inferred from the ratio of the power of this model to the \LCDM spectrum at large scales (0.806 compared to 0.794). We chose to use the $\sigma_8$ value inferred from this ratio in our \halofit and \hmcode calculations for our results, which ensures that the large-scale response in Fig.~\ref{fig:codecs} is accurate.

%Halo model calculation
The results of our halo-model calculation are shown in Fig.~\ref{fig:codecs}, where we show the response from the \codecs simulations together with that predicted by \halofit and the \cite{Mead2015b} halo model. One can see that both models perform similarly, but that the halo model has improved accuracy when modelling the bump that starts around $k\simeq 0.2\iMpc$. In the halo model, this bump is power due to halo shot noise, and should be captured by both methods through their dependence on $\sigma(R)$ under the assumption of mass function universality. The mass function in the \codecs simulations has been investigated by \cite{Cui2012}, who show that at $z=0$ and $1$ deviations from universality never exceed 10 per cent apart from for the highest mass haloes, where they reach a maximum of 20 per cent, this is despite the great differences in growth \emph{history}; this is in accordance with \cite{Press1974} theory which suggests that the linear power spectrum determines the mass function. This explains the relatively good performance of both the halo model and \halofit for $k<1\iMpc$. However, both models fail to predict the power for $k>1\iMpc$, which must be a consequence of neither model correctly accounting for the CDM--DE energy transfer, and the effect of this on the internal structure of the halo. Within the framework of \halofit it is difficult to improve matters, but with the halo model we still retain knowledge of the halo profiles and so can modify halo internal structure to account for the coupling.

\begin{table}
\begin{center}
\caption{Values of the amplitude (A) of the $c(M)$ relation (equation \ref{eq:bullock_cm}) that best-fitting power spectrum data from the \codecs simulations. The standard \citeauthor{Mead2015b} (\citeyear{Mead2015b}) value calibrated to $w$CDM is $A=3.13$ independent of redshift.}
\begin{tabular}{c c c}
\hline
Model & A ($z=0$) & A ($z=1$) \\
\hline
\LCDM & 3.13 & 3.13 \\
EXP001 & 3.02 & 3.07 \\
EXP002 & 2.78 & 2.90 \\
EXP003 & 2.48 & 2.55 \\
EXP008e3 & 2.59 & 2.85 \\
SUGRA003 & 6.59 & 5.31\\
\hline
\end{tabular}
\end{center}
\label{tab:codecs_concentration}
\end{table}

%Fitting c(M)
Rather than attempting to calculate the necessary changes to halo profiles using theoretical arguments, we opt to fit the amplitude of the concentration--mass relation ($A$ in equation~\ref{eq:bullock_cm}) within the halo model to the power spectrum of each \codecs model at each $z$. The result of this fit is shown as the solid-black line in Fig.~\ref{fig:codecs}, where we see that a few per cent level accuracy can be achieved for each model. The best-fitting values of $A$ for each model and $z$ are given in Table~\ref{tab:codecs_concentration}. Without this amplitude shift, the (\citealt{Bullock2001} with \citealt{Dolag2004} augmentation) relation predicts similar $c(M)$ for all models, but there are some small differences due to the different growth history in each model. However, these differences do not exceed $25$ per cent for any model at any halo mass. It is expected that the \cite{Bullock2001} concentrations would be appropriate for a minimally coupled quintessence model, but they do not have knowledge of the $\beta$-coupling, and thus will fail for the \codecs models. Pleasingly, the required amplitude shifts in the concentration--mass relation necessary to provide accurate halo-model power spectra reflect concentration differences measured in haloes from the \codecs simulations. For example, \cite{Cui2012} and \cite{Giocoli2013} show that the SUGRA003 model has halo concentrations that are enhanced by a factor of $\simeq 2$, while concentrations are similar to, or slightly less than, the \LCDM value in the other models. This trend is exactly what we need to remedy our halo model results. If some way was found to predict $c(M)$ from first principles then our results indicate that the halo model would deliver accurate power for $k<10\iMpc$. Using spherical model arguments \citep{Wintergerst2010} it may be possible to predict the $c(M)$ change via changes in $\dc$ (which appears in the \citealt{Bullock2001} relations) and $\Dv$, but we did not pursue this. 

%Summary
Alternative approaches have been investigated for predicting the non-linear spectrum in the \codecs models. \cite{Baldi2012b} investigated \halofit predictions and show the same results as can be seen in Fig. \ref{fig:codecs}. \cite{Casas2016} produced accurate fits to the \codecs simulation response using five fitting parameters for each model, and then forecast the constraining power of future surveys. Although our fits are less accurate, we suggest that using the halo model is preferable in general because it has the potential to capture the cosmology dependence of the response (\eg the dependence on $\Om$). However, this would require a physically motivated model for the change in halo internal structure, which we do not have.

\vspace{1cm}
\subsection{Massive neutrinos}
\label{sec:massive-nu}

%Introduction
The cosmological background number density of standard model, active neutrinos can be predicted exactly from thermal equilibrium arguments, and is set when the neutrinos decouple from other matter in the early Universe. Therefore fixing the mass of each neutrino fixes the physical cosmological neutrino density parameter:
\begin{equation}
\Onu h^2\simeq\frac{\sum_\nu m_\nu}{94.1\eV}\ .
\end{equation}
Three neutrino species appear in the standard model and have been measured by particle physics experiments. Neutrinos contribute to the radiation density in the early Universe and therefore have observable consequences for structure formation because they prolong the radiation epoch and this delays the time at which matter perturbations can start to collapse. The number of extra radiation species can therefore be measured through cosmological observations and these measurements are also consistent with there being three neutrino species \citep[\eg][]{PlanckXIII2015}.

%Neutrino mass
Flavour oscillations point definitively to a non-zero mass for at least two of the three neutrino mass eigenstates \citep{Fukuda1998,SNO2002}, but oscillation experiments only measure mass--squared differences between eigenstates, and not the absolute masses. For standard-model active neutrinos, this places a minimum combined mass for all neutrino species of $m_\nu \gtsim 0.06\eV$ ($0.1\eV$) for the normal (inverted) hierarchy \citep{Otten2008}. Constraints from $\beta$-decay experiments place an upper limit of $m_{\nu,\mathrm{e}}<2.2\eV$ for the electron neutrino \citep{Lobashev2003}. From cosmology: at the upper end, simply requiring the Universe not to be closed places a limit of $\sum_\nu m_\nu\ltsim94.1 h^2\eV$ \citep{Gershtein1966} while probes of cosmological structure place limits of $m_\nu<0.23\eV$ \citep{PlanckXIII2015} and even tentative detections of a non-zero mass \citep{SPT2014} from the CMB in combination with other cosmological data. More recently, limits of $m_\nu<0.12\eV$ have been obtained from the Lyman-$\alpha$ forest \citep{Palanque-Delabrouille2015} in combination with CMB data. 

%Neutrinos in cosmology - HDM
Given the cosmological constraints, it is known that neutrinos cannot make up the entirety of the DM, but they \emph{will} make up a small mass fraction. These neutrinos are relativistic when they decouple, and only become non-relativistic after recombination. Given the number of neutrinos that are known to exist, this places a lower limit on the neutrino density parameter of $\Onu\gtsim0.001$. We define $f_\nu=\Onu/\Om$ and $\Om$ as the \emph{total} matter density parameter $\Om=\Oc+\Ob+\Onu$ with $\Oc$ being the CDM density. Cosmologically viable neutrinos are very light, and thus have large free-streaming lengths, and are considered a species of hot DM. This washes perturbations out in the neutrino component on small scales, below free-streaming, and this has a back reaction effect on the full matter distribution. For a fixed matter density clustering is suppressed at linear order below the free-streaming scale because a fraction of matter below this scale is smoothly distributed. If $\Om=1$ and $f_\nu\ll 1$ then the growth factor changes from $\propto a$ to $\propto a^{1-3f_\nu/5}$. Despite small values of $f_\nu$ allowed by observations, having a fraction of mass in neutrinos suppresses linear power by a surprisingly large amount, $\sim 8f_\nu$ at fixed $\Om$, because the suppression is an integrated effect from the time at which perturbations start to develop. The mildly non-linear effect of neutrinos can be investigated using perturbation theory \new{\citep[\eg][]{Wong2008,Pietroni2008,Saito2008,Saito2009,Lesgourgues2009,Upadhye2014,Peloso2015}}
, or via simulations, which show that the non-linear power is suppressed by a greater amount than the linear \new{prediction for the suppression} \new{\citep[\eg][]{Brandbyge2010a,Brandbyge2010b,Agarwal2011,Bird2012,Massara2014}}. Thus examining structure via weak lensing is a promising methods for determining the mass of the neutrino and, in principle, the non-linear matter distribution is more discriminating than the linear. The sensitivity to non-linear scales makes weak lensing an ideal tool to probe the neutrino mass \cite[\eg][]{Kitching2008,Ichiki2009} and limits have already been placed using CFHTLenS data by \cite{Battye2014}, \cite{MacCrann2015} and \cite{Harnois-Deraps2015a}.

\begin{table}
\begin{center}
\caption{Parameters of the massive neutrino simulations from \citecaption{Massara2014} that were used in this paper. $\Om=0.3175$, $\Ob=0.049$, $\Ov=1-\Om$, $h=0.671$ and $n_\mathrm{s}=0.962$ are fixed for each model, but as the total neutrino mass is increased the CDM density, $\Oc$, decreases. Transfer functions were generated using \camb and each of the three neutrinos was assigned a third of the total neutrino mass.}
\begin{tabular}{c c c c c c c}
\hline
Model & $\sum_\nu m_\nu/\eV$ & $\Om$ & $\Onu$ & $\Oc$ &$f_\nu$ & $\sigma_8$\\
\hline
\LCDM & 0.00 & 0.3175 & 0.0000 & 0.2685 & 0.000 & 0.834\\
M015   & 0.15 & 0.3175 & 0.0035 & 0.2650 & 0.011 & 0.800\\
M030   & 0.30 & 0.3175 & 0.0071 & 0.2614 & 0.022 & 0.763\\
M060   & 0.60 & 0.3175 & 0.0142 & 0.2543 & 0.044 & 0.693\\
\hline
\end{tabular}
\label{tab:nu}
\end{center}
\end{table}

%Running of massive-nu simulations
Simulating massive neutrinos involves some subtleties: normally in gravity-only simulations, one considers the clustering of `all matter' and does not differentiate between matter species. This is despite the great differences in transfer functions for baryons and CDM at the high redshifts at which simulations typically begin. These differences arise due to different physical processes acting on different \new{species} prior to recombination. This is accounted for by taking the linear-theory all-matter power spectrum at $z=0$, when the transfer functions of baryons and CDM \emph{are} nearly identical, and evolving this back to the simulation start redshift using a scale-independent growth function that is appropriate for `all matter'. Whilst this ensures that the correct $z=0$ linear matter power is evolved, it means that the simulation will not be an accurate reflection of the real Universe at high $z$ when baryons and CDM are strongly differentiated. In contrast, when massive neutrinos are simulated one \emph{must} consider the spectra of cold and hot matter separately, because their linear clustering is very different at $z=0$. Neutrinos have been incorporated into simulations in a variety of ways: the simplest possible method is to treat neutrinos only via their effects on the background expansion, assuming they are always in the linear regime \citep[\eg][]{Agarwal2011}. This is justified because the large free-streaming lengths of viable neutrinos ensure they are distributed more linearly and with a more Gaussian distribution than the CDM. A more accurate method is to evolve neutrinos in tandem with CDM particles, but using neutrino linear equations and simulating them on a mesh rather than as a separate particle species \citep[\eg][]{Brandbyge2009}. This means that any non-linear structure in the neutrinos is missing, which makes a small impact at the level of the matter power spectrum. The final, and potentially most accurate, option is to include the neutrinos as a separate particle species in the simulation \citep[\eg][]{Brandbyge2008,Viel2010}, thus capturing all aspects of their non-linear evolution and the back-reaction of this on the CDM. These methods have been compared by \eg \cite{Agarwal2011}, \cite{Bird2012} and \cite{Upadhye2014} and there is some disagreement about the accuracy of each method, particularly because particle neutrinos are difficult to implement in practice due to the large neutrino thermal velocities at high redshifts.

%Scale-dependent growth and sigma_cb
A consequence of the discussion in the above paragraph is that we must account for the different scale-dependent linear growth functions for cold and hot matter in our halo model. In weak lensing, one is interested in the power spectrum of all matter, but from a clustering point of view it is the cold matter that is of prime importance in the formation of dense virialized structures. This means that in the halo model we should calculate $\sigma(R)$, which regulates clustering, using the cold matter component only (\ie CDM and baryons; $\sigma_\mathrm{cb}$) and ignoring the contribution from hot matter. Using this definition of $\sigma(R)$ was shown to produce more `universal' (function of $\sigma$ only) mass functions in $\nu$\LCDM models by \cite{Brandbyge2010a} and \cite{Castorina2014}, more universal halo bias by \cite{Villaescusa-Navarro2014} and a more accurate halo-model power spectrum response by \cite{Massara2014}. Although this approach ignores neutrino non-linearity, it should be a good approximation to describe the suppression in the non-linear clustering of cold matter in response to the neutrinos. In reality, we expect the neutrinos to start falling into the potential wells of the cold matter once they become non-relativistic and to eventually form neutrino haloes around those of CDM \new{\citep[\eg][]{Ringwald2004,Abazajian2005}}; but this is ignored in our modelling. Note that the conservation of the neutrino phase-space density prevents these neutrino haloes from becoming as dense as their CDM counterparts \citep{Tremaine1979}.

%This approximation was investigated in \cite{Angulo2013a} where it was shown that properly including distinct transfer functions for DM and baryons leads only to small differences in the eventual measured non-linear matter power spectrum -- below one per cent at late times around the BAO scale at $z=0$.

\begin{figure*}
\begin{center}
\includegraphics[angle=270,width=16cm]{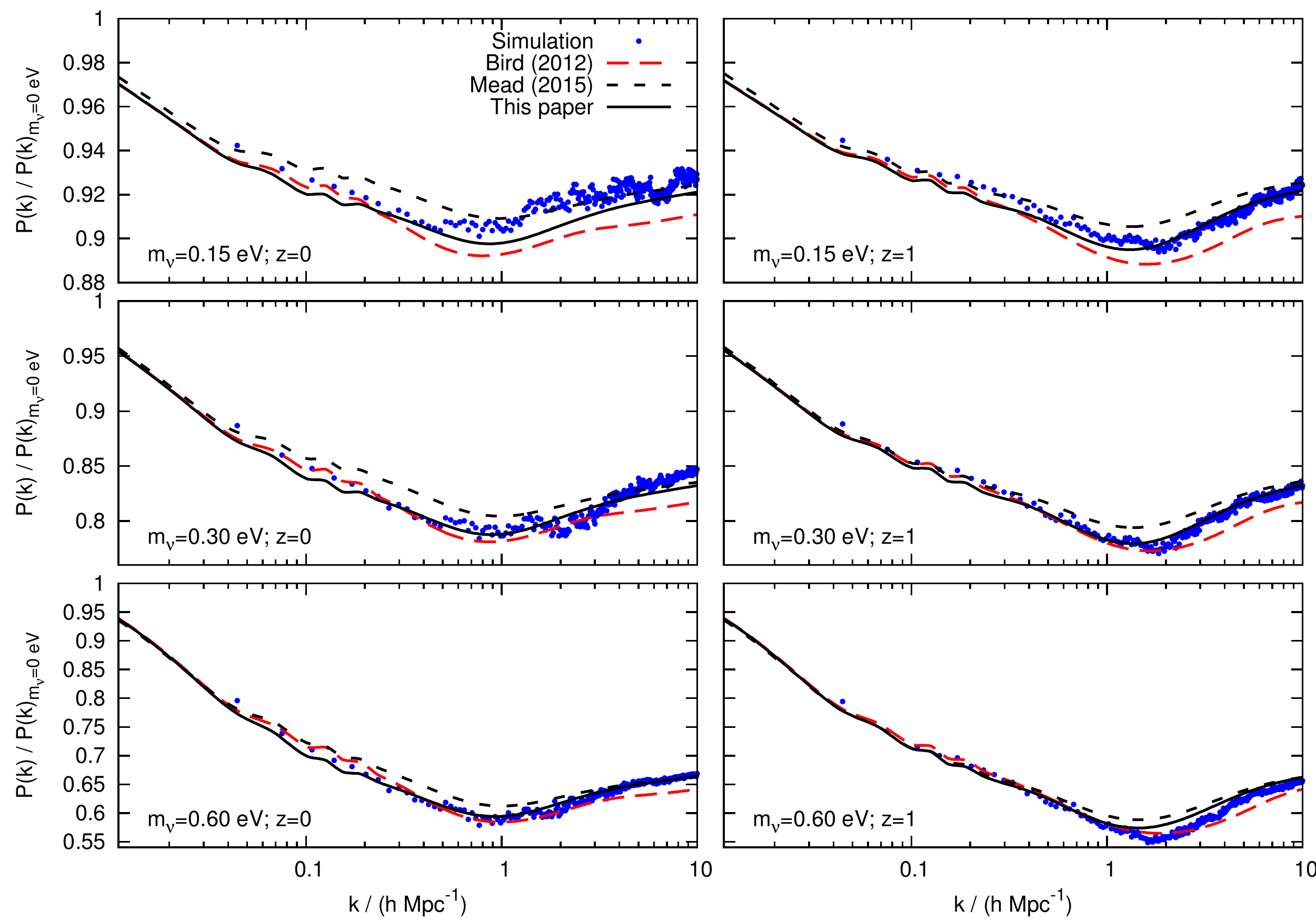}
\end{center}
\caption{A comparison of the power spectrum response for massive-$\nu$ models with three degenerate neutrinos with total mass $0.15\eV$ (top) $0.3\eV$ (middle) and $0.6\eV$ (bottom) compared to an equivalent \LCDM model (with $\Om$ fixed between models, rather than $\Oc$) at $z=0$ (left-hand column) and $1$ (right-hand column). We show the response from the simulations of \citeauthor{Massara2014} (\citeyear{Massara2014}; blue points), that from the \citecaption{Bird2012} version of \halofit (long-dashed; red) and that from the \citecaption{Mead2015b} halo model (short-dashed; black). We see that all models of the response are in broad agreement, but that the halo model of \citecaption{Mead2015b} mis-predicts the degree of quasi-linear damping. Our updated version of the halo model, with tuned parameters (solid; black), matches the simulations at the few per cent level across the full range of scales shown, with the eventual agreement being similar to, but slightly better than, that of \citecaption{Bird2012}.}
\label{fig:nu}
\end{figure*}

\begin{figure*}
\begin{center}
\includegraphics[angle=270,width=17cm]{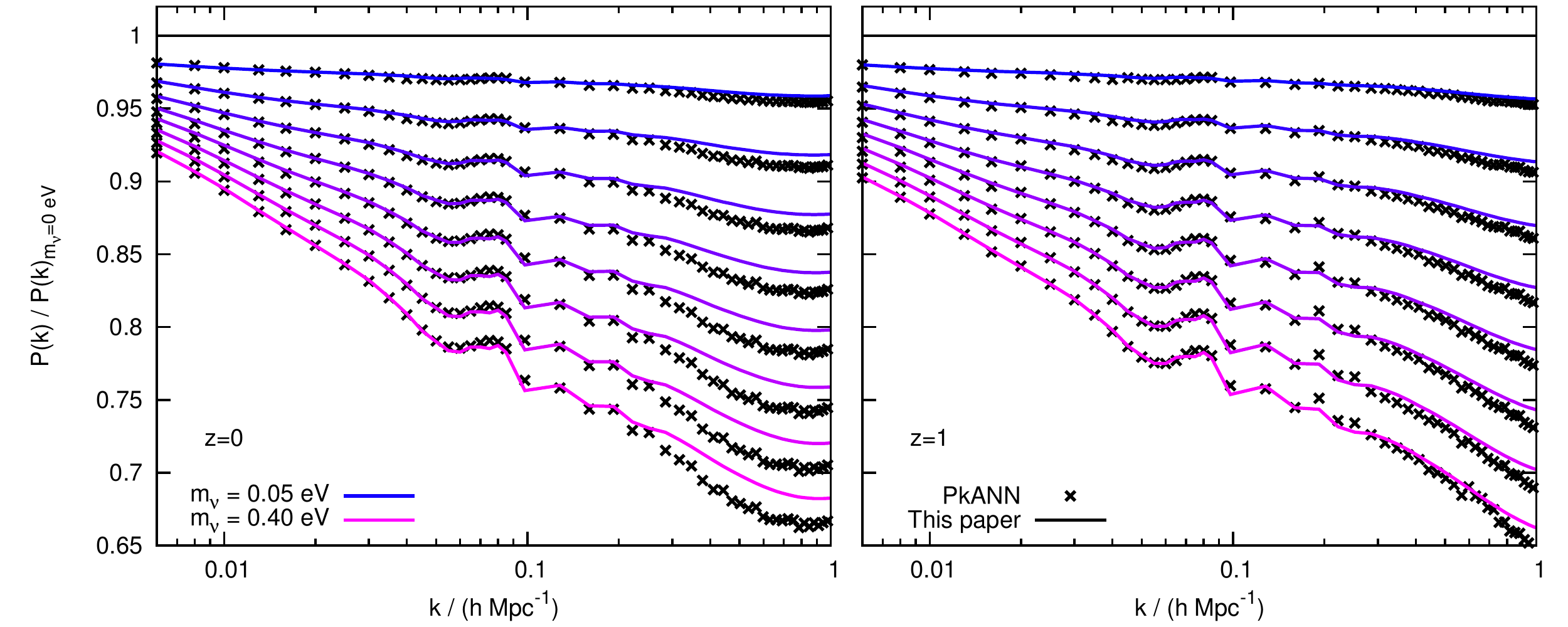}
\end{center}
\caption{A comparison of the power spectrum response from the halo model approach of this paper (solid lines) compared to the \pkann emulator (crosses) of \citecaption{Agarwal2014} at $z=0$ (left) and $1$ (right). We show the response as the total neutrino mass (three degenerate species) is varied linearly between $0.05$ (blue; top curve) to $0.4\eV$ (pink; bottom curve) while $\Om h^2$ and $\Ob h^2$ are held fixed and $h$ is adjusted to match the acoustic scale. We simultaneously vary $\sigma_8$ between 0.8 (blue) and 0.7 (pink) to keep the large scale portion of the power spectrum similar (note well that this is not perfect). We see few per cent level agreement across most of the range of scales up to $k=1\iMpc$. The larger differences around $k\simeq1\iMpc$ at $z=0$ are consistent with the differences in simulation scheme between the simulations used for \pkann and those to which our model was tuned.}
\label{fig:pkann}
\end{figure*}

%Discussion of the figure
We compare halo-model predictions from the \cite{Mead2015b} model to the massive-$\nu$ simulations presented in \cite{Massara2014}, which consider neutrinos with $m_\nu=0.15$, $0.3$ and $0.6\eV$ in a box of $L=200\Mpc$ with $N=512^3$ of both CDM and neutrino particles and are detailed in Table~\ref{tab:nu} Both `all matter' and baryon densities are held fixed as the neutrino mass is varied, so increasing $m_\nu$ consequently decreases $\Oc$. The neutrino mass is taken to be evenly distributed between three degenerate species, even though this is in conflict with oscillation experiments for low $m_\nu$. Using the prescription for the halo model advocated by \citeauthor{Massara2014} (\citeyear{Massara2014}, \ie using $\sigma_\mathrm{cb}$ for clustering calculations), we present our results in Fig. \ref{fig:nu}. We show the response from simulations together with that from the halo model of \cite{Mead2015b} and from the fitting formula of \cite{Bird2012}, which is an update of the \cite{Smith2003} version of \halofit\footnote{Note that we show the response from the published \cite{Bird2012} appendage to the original \cite{Smith2003} \halofit. This is \emph{not} the version currently implemented in \camb, which contains some unpublished corrections.}. We see that all models do a reasonable job of predicting the suppression of power that peaks around $k=1\iMpc$, which is caused by the massive neutrinos suppressing CDM clustering. The \cite{Bird2012} model is accurate at the $3$ per cent level, but seems to over predict the power suppression at $k>1\iMpc$, whereas \cite{Mead2015b} halo model does well for $k>1\iMpc$, but under predicts the magnitude of the quasi-linear ($k\sim 0.1\iMpc$) damping.

%Improvements to the halo model for damping region
To improve the halo-model predictions, we note that the magnitude of quasi-linear damping is governed by $f$ in equation~(\ref{eq:2haloterm}), which depends on $\sigma_8(z)$ in \cite{Mead2015b}. In massive neutrino models, clustering is suppressed and $\sigma_8$ drops quite drastically as the neutrino mass is increased (see Table~\ref{tab:nu}), which in turn changes $f$ and causes the under prediction of damping. We remedy this by re-parameterizing $f$ in terms of $\sigma_\mathrm{d}(R)$ (\ie -- the standard deviation in the linear displacement field convolved with top-hat filter of radius $R$, which is less influenced by small scales than $\sigma(R)$) where we found good matches to \emu power spectra using $R=100 \Mpc$. The updated form of $f$ is given in Table~$\ref{tab:fit_params}$. In order to maintain a good fit to the \emu simulations as obtained in \cite{Mead2015b} we simultaneously refit the coefficients of $f$ and the quasi-linear $\alpha$ term in equation~(\ref{eq:meadfit}); updated values are given in Table~$\ref{tab:fit_params}$. This actually makes a small improvement to the quality of the fit to the \emu simulations that was presented in fig. 2 in \cite{Mead2015b}.

%Fitting of Dv and dc
To further improve predictions we use the spherical model of non-linear structure formation  to calculate values for the linear-collapse density ($\dc$) and virialized over-density ($\Dv$) for an isolated top-hat density perturbation. These values will change in massive neutrino models because some fraction of the matter is unclustered. The spherical model for massive-$\nu$ cosmologies has been considered in detail in \cite{Ichiki2012} and \cite{LoVerde2014} but we consider a simpler model where we work in the limit that neutrinos are completely unclustered, and therefore only contribute to the background expansion. In this case, if $\Om=1$, we find a good match to spherical model results with
\begin{equation}
\eqalign{
\dc&\simeq 1.686\times (1-0.041f_\nu)\ , \cr
\Dv&\simeq 178\times (1+0.763 f_\nu)\ .
}
\label{eq:spherical_nu}
\end{equation}
We also found that, when including $\Lambda$, the \emph{deviation} from the $\Lambda$CDM prediction was only weakly dependent on $\Lambda$ (\ie the $f_\nu$ dependence in equations~\ref{eq:spherical_nu} holds). Taking these functional forms for $\dc$ and $\Dv$ as inspiration, we fit for the coefficients of the dependence on $f_\nu$ to the simulation data of \cite{Massara2014} at $z=0$ and $1$ for $m_\nu=0.15$, $0.3$ and $0.6\eV$. The best fitting values were found to be
\begin{equation}
\eqalign{
\dc&\propto 1+0.262 f_\nu\ , \cr%[1+0.0123\log_{10}\Om(z)]\ , \cr
\Dv&\propto 1+0.916f_\nu\ . 
}
\label{eq:spherical_nu_fit}
\end{equation}
We note that the best fitting dependence on $\dc$ is opposite to, and much stronger than, the spherical model dependence, but that of $\Dv$ is of a similar magnitude and sign. Our fitted halo model is shown as the solid black line in Fig.~\ref{fig:nu} where we see a per cent level match to simulations across all scales and a small improvement over the fitting formula of \cite{Bird2012}. It should be noted that this has been achieved using only two free parameters to fit a range of neutrino masses at two different redshifts. It is possible that some of the fitting of $\dc$ and $\Dv$ accounts for neutrino non-linearity, which we have not accounted for explicitly in our modelling. In this future this could be accounted for using the halo modelling of \cite{Abazajian2005}, although their results suggest that the effects of neutrino non-linearity are around 1 per cent for the neutrino masses we consider here. We also note that, prior to fitting $\dc$ and $\Dv$, our model is already a rather good match to the simulation data. This is not shown in Fig.~\ref{fig:nu}, but the match is of similar accuracy to the \cite{Bird2012} model.

In Fig.~\ref{fig:pkann}, we show a comparison of the response of our updated halo-model to the matter power spectrum prediction emulator of \pkann \citep{Agarwal2012,Agarwal2014} where we vary the neutrino mass within the range $0$-$0.4\eV$ and simultaneously vary $\sigma_8$ from 0.8 to 0.7 to keep the large-scale power similar. The simulations that were used to create \pkann treat the neutrinos as a linear component and they only affect cold matter via their effects on the background. Note that $h$ is \emph{not} a free parameter in \pkann and is set so as to match the combined Wilkinson Microwave Anisotropy Probe 7 year and BAO acoustic scale results \citep{Komatsu2011}. Since the physical densities $\Omega_\mathrm{c}h^2$ and $\Omega_\mathrm{b}h^2$ are also fixed this implies a change in $\Om$, $\Oc$ and $\Ob$. We see a good (few per cent level) agreement for all neutrino masses up to the smallest scale output by \pkann ($k\simeq1\iMpc$) at $z=0$ and $1$. However, there is a slight mismatch between the calibrated halo model and \pkann for the neutrino induced damping around $k=1\iMpc$ at $z=0$, which is consistent with the differing simulations schemes used by \cite{Massara2014} and \cite{Agarwal2014}. \cite{Bird2012} show that linear neutrino simulations over predict the magnitude of the neutrino-induced suppression around $k=1\iMpc$ when compared to Fourier-space methods. The linear scheme will be more accurate for lower neutrino masses and it is here that we see the best overall agreement between \pkann and the halo model. Note also that the simulations run for the \pkann project incorporate some hydrodynamics, and this may conceivably account for the poorer agreement around $k=1\iMpc$ at $z=0$. As we compare our results at the level of the response, not the absolute power, hydrodynamics may well cancel out, unlike differences from the neutrino simulation method. %Indeed, our results are more discrepant if we compare the absolute value predicted by \meadfit to that of \pkann, but are still accurate at the 4 per cent level for all neutrino masses.

\subsection{Chameleon screening}
\label{sec:chameleon}

%Introduction to chameleon screened models
In this section, we work with the \cite{Hu2007a} $f(R)$ model that exhibits the chameleon screening mechanism. $f(R)$ models \citep{Buchdal1970,Nojiri2003,Carroll2005} are derived from a modified Einstein--Hilbert action, in which a general $f(R)$ is added to the standard linear $R$ term. In this work, we use the high curvature limit of the \cite{Hu2007b} $f(R)$ function that is widely deployed throughout the literature:
\begin{equation}
f(R)=-2\Lambda-\bar{R}_0\frac{f_{R0}}{n}\left(\frac{\bar{R}_0}{R}\right)^n\ ,
\label{eq:hu_sawicki_approx}
\end{equation}
where $f_{R0}$ and $n$ are the model parameters and $\bar{R}_0$ is the background value of the Ricci scalar, $R$, measured today. Here, $f(R)$ has the form of a (cosmological) constant plus a modification term and the mechanism for accelerated expansion ($-2\Lambda$) is entirely divorced from that which directly modifies gravitational forces. We consider only models with $f_{R0}<0$ and $n>0$, and we work in the limit where $|f_{R0}|\ll 1$ (which covers values that are interesting observationally) such that the inverse $R$ term is negligible when considering the evolution of the background, which means that the background expansion is that of a \LCDM model. This has the advantage of allowing us to study the modification in isolation from effects that may arise from a different expansion history. Throughout this section, we use units such that $c=1$.

%Maths and linear theory
The derivative of $f(R)\equiv f_R$ can be considered a new scalar field, related to $R$ via:
\begin{equation}
f_R=f_{R0}\left(\frac{\bar{R}_0}{R}\right)^{n+1}\ .
\label{eq:f_R_HS07}
\end{equation}
At the background level
\begin{equation}
\bar R=3H_0^2(\Om a^{-3} +4\Ov)\ ,
\end{equation}
and
\begin{equation}
\bar{f}_R(a)=f_{R0}\left(\frac{1+4\Ov/\Om}{a^{-3}+4\Ov/\Om}\right)^{n+1}\ .
\label{eq:background_fr}
\end{equation}
In the quasi-static limit the equation that governs departures of $f_R$ from the background is:
\begin{equation}
\frac{1}{a^2}\nabla^2\delta f_R=\frac{1}{3}\delta R-\frac{8\pi G}{3}\bar{\rho}_\mathrm{m}\delta\ ,
\label{eq:quasistatic_fr}
\end{equation}
where, $\delta f_R\equiv f_R-\bar{f}_R$, $\delta R\equiv R-\bar{R}$ and the Laplacian is comoving and we have \emph{not} assumed that $|\delta f_R|$ is small in comparison with $|f_{R0}|$. Non-relativistic particles in an $f(R)$ model feel a modified acceleration compared to standard gravity counterparts. This can be seen most easily via the perturbed metric in flat space:
\begin{equation}
\mathrm{d}s^2=(1+2\Psi)\,\mathrm{d}t^2-a^2(t)(1-2\Phi)\,\mathrm{d}\mathbf{x}^2\ ,
\label{eq:weak_field_metric}
\end{equation}
from which equations for the time-gravitational potential $\Psi$ and space-gravitational potential $\Phi$ can be derived:
\begin{equation}
\frac{1}{a^2}\nabla^2\Psi=\frac{16\pi G}{3}\bar{\rho}_\mathrm{m}\delta-\frac{1}{6}\delta R\ ,
\label{eq:psi}
\end{equation}
\begin{equation}
\frac{1}{a^2}\nabla^2\Phi=\frac{8\pi G}{3}\bar{\rho}_\mathrm{m}\delta+\frac{1}{6}\delta R\ .
\label{eq:phi}
\end{equation}
Non-relativistic particles are accelerated by the $\Psi$ potential, and are thus affected by the $\delta f_R$ field. Photon trajectories (and thus lensing) are governed by $\Psi+\Phi$, which is unchanged from that in standard gravity as long as $|f_{R0}|\ll 1$. If $\delta f_R$ is small then we may linearize:
\begin{equation}
\delta f_R\simeq \left.\frac{\mathrm{d}f_R}{\mathrm{d}R}\right|_{\bar R}\delta R\equiv\frac{1}{3}\lambda^2\delta R\ ,
\end{equation}
where $\lambda$ is known as the `Compton wavelength' and is the scale at which the linear modification becomes active. In the model considered here
\begin{equation}
\lambda^2=-3(n+1)\frac{f_{R0}}{\bar{R}_0}\left(\frac{\bar{R}_0}{\bar{R}}\right)^{n+2}\ .
\label{eq:compton}
\end{equation}
Larger $|f_{R0}|$ values mean the modification is felt at larger scales. Linearizing the equation for $\Psi_k$ allows us to see that the sub-horizon growth of linear matter perturbations is scale dependent:
\begin{equation}
\ddot\delta_k+2H\dot\delta_k=\frac{3}{2}H^2\Omega_\mathrm{m}(a)\left[1+\frac{1}{3}\left(\frac{1}{1+a^2/\lambda^2 k^2}\right)\right]\delta_k\ ,
\label{eq:perturbations}
\end{equation}
where the over-dots denote physical time derivatives. For linear scales much smaller than the Compton wavelength gravity is enhanced by a factor $4/3$, whereas gravity is standard on large scales.

%Introduction to chameleon screening
A remarkable feature of $f(R)$ models is that they can screen the effect of the $4/3$ modification in massive haloes and that this screening exhibits itself naturally, without it having to be introduced by hand. This screening allows stringent tests of gravity within the Solar system to remain satisfied (see \eg \citealt{Will2006}), while modifying gravity on larger scales. It was shown that $f(R)$ models can exhibit the chameleon mechanism by \cite{Li2007}, \cite{Hu2007b} and \cite{Brax2008}. If $\nabla^2 \delta f_R=0$ then
\begin{equation}
\frac{1}{3}\delta R=\frac{8\pi G}{3}\bar{\rho}_\mathrm{m}\delta\ ,
\end{equation}
and gravitational forces are restored to the standard. Determining when this condition is satisfied in a cosmological context can be achieved using simulations (\citealt{Oyaizu2008a,Zhao2011,Li2012a,Puchwein2013,Llinares2014}; summarized by \citealt{Winther2015b}) or via direct calculations in idealized situations with symmetry properties \citep[\eg][]{Hu2007b,Schmidt2010a,Lombriser2012c}. These calculations show that the screening depends primarily on halo mass, with screening extending to lower halo masses for smaller $|f_{R0}|$ values. The fact that the Milky Way is screened, and has a potential $\mathcal{O}(10^{-6})$, can be used to place limits of $|f_{R0}|\ltsim 10^{-6}$ \citep{Hu2007b} on these models.

%Constraints from clusters yield $|f_{R0}|\ltsim 10^{-4}$ \citep{Schmidt2009,Ferraro2011,Lombriser2012a,Lombriser2012b}.

%Observational constraints
Alternatively, limits on chameleon models can be placed by searching for differences between samples of similar objects in screened and unscreened environments (\eg dwarf galaxies -- \citealt{Jain2013}; \citealt{Vikram2013}) and constraints of $|f_{R0}|\ltsim 10^{-7}$ are obtained. Independent constraints can be placed from large-scale structure measurements -- particularly from the abundance of clusters, which increases in these models due to the enhanced large-scale gravity. \cite{Cataneo2015} report constraints of $|f_{R0}|\ltsim 10^{-5}$ from cluster abundance and \cite{Terukina2014} use the difference between hydrostatic and lensing masses (\ie, the same mass cluster will have a higher X-ray temperature if it is unscreened) to infer constraints of $|f_{R0}|< 6\times 10^{-5}$. \cite{Li2016} use cluster gas fractions to place limits of $|f_{R0}|< 5\times 10^{-5}$. Cosmologically, \cite{Dossett2014} used redshift space distortions in the WiggleZ survey to place the limit of $|f_{R0}|\ltsim 10^{-5}$. \cite{Harnois-Deraps2015c} rules out $|f_{R0}|=10^{-4}$ using \cfhtlens data while simultaneously accounting for massive neutrinos and baryonic feedback (which can partially compensate for the enhanced gravity; see Section~\ref{sec:degeneracy}). While astrophysical constraints are currently tighter than cosmological this may change in the future. \cite*{Lombriser2015} show that the modification signal in the power spectrum can be enhanced in simulations by using the clipping technique \citep{Simpson2011,Simpson2013b} to `unscreen' the modification; removing high-density peaks in the density field exposes more of the unscreened matter to the power spectrum. The clipping technique has recently been applied to lensing simulations by \cite{Simpson2016}, paving the way for a direct application of unscreening to lensing data.

%Lucas' stuff
We note that it is theoretically feasible for the $f_R$ field may couple only to DM (if the model is couched as a scalar field with non-universal couplings in the \emph{Einstein} frame), and that this would invalidate Solar system and Galactic constraints, meaning that the model may \emph{only} be constrained on cluster or cosmological scales. If Solar system and baryonic constraints are excluded then a conservative bound on current limits is $|f_{R0}|\ltsim 10^{-5}$ otherwise this limit is more like $|f_{R0}|\ltsim 10^{-7}$. Note that all constraints quoted are $2\sigma$ for $n=1$ \cite{Hu2007b} models; constraints on $|f_{R0}|$ are degraded if $n>1$ because these models transition to mimic $\Lambda$CDM more quickly in the recent past and the linear enhancement in power is therefore less strong because it is a cumulative effect.

%Halo profiles unchanged
One might expect the enhanced gravity to change the halo profile, but it has been shown \citep[\eg][]{Lombriser2012c} that haloes in $f(R)$ simulations can be well described by NFW profiles with close to standard concentration--mass relations (although there seems to be a small transition in amplitude at low mass -- fig.4 of \citealt{Shi2015}). However, as would be expected from \cite{Press1974} theory, the mass function is enhanced because $\sigma(R)$ is boosted due to the enhanced growth of linear perturbations 

%\cite{Lombriser2014b}

\begin{table}
\centering
\caption{Simulations of standard gravity (GR) and $f(R)$ (F4, F5, F6) models from \citeauthor{Li2013} (\citeyear{Li2013}) that are used in this paper. The cosmological parameters are $h = 0.73$, $\Om =0.24$, $\Ob = 0.042$, $\Ov = 1-\Om$, $n_\mathrm{s} = 0.958$ and $\sigma_8 = 0.77$. All $f(R)$ models have $n=1$ but differing values of $f_{R0}$. Simulations begin at $z_\mathrm{i}=49$ in a cube of side $512\Mpc$ from exactly the same initial conditions, but growth is enhanced in models with larger $|f_{R0}|$ values. The inverse Compton scale (equation~\ref{eq:compton}) at $z=0$ is also shown for each model, and indicates the $k$-scale at which the modification is active.}
\begin{tabular}{c c c c c}
\hline
Simulation & $n$ & $f_{R0}$ & $\sigma_8(z=0)$ & $1/\lambda$ \\ [0.5ex] 
\hline
GR & -- & -- & 0.770 & --            \\
F6 & 1 & $-10^{-6}$ & 0.784 & $0.427\iMpc$ \\
F5 & 1 & $-10^{-5}$ & 0.824 & $0.135\iMpc$ \\
F4 & 1 & $-10^{-4}$ & 0.887 & $0.043\iMpc$ \\
\hline
\end{tabular}
\label{tab:fR}
\end{table}

\begin{figure*}
\begin{center}
\includegraphics[angle=270,width=17cm]{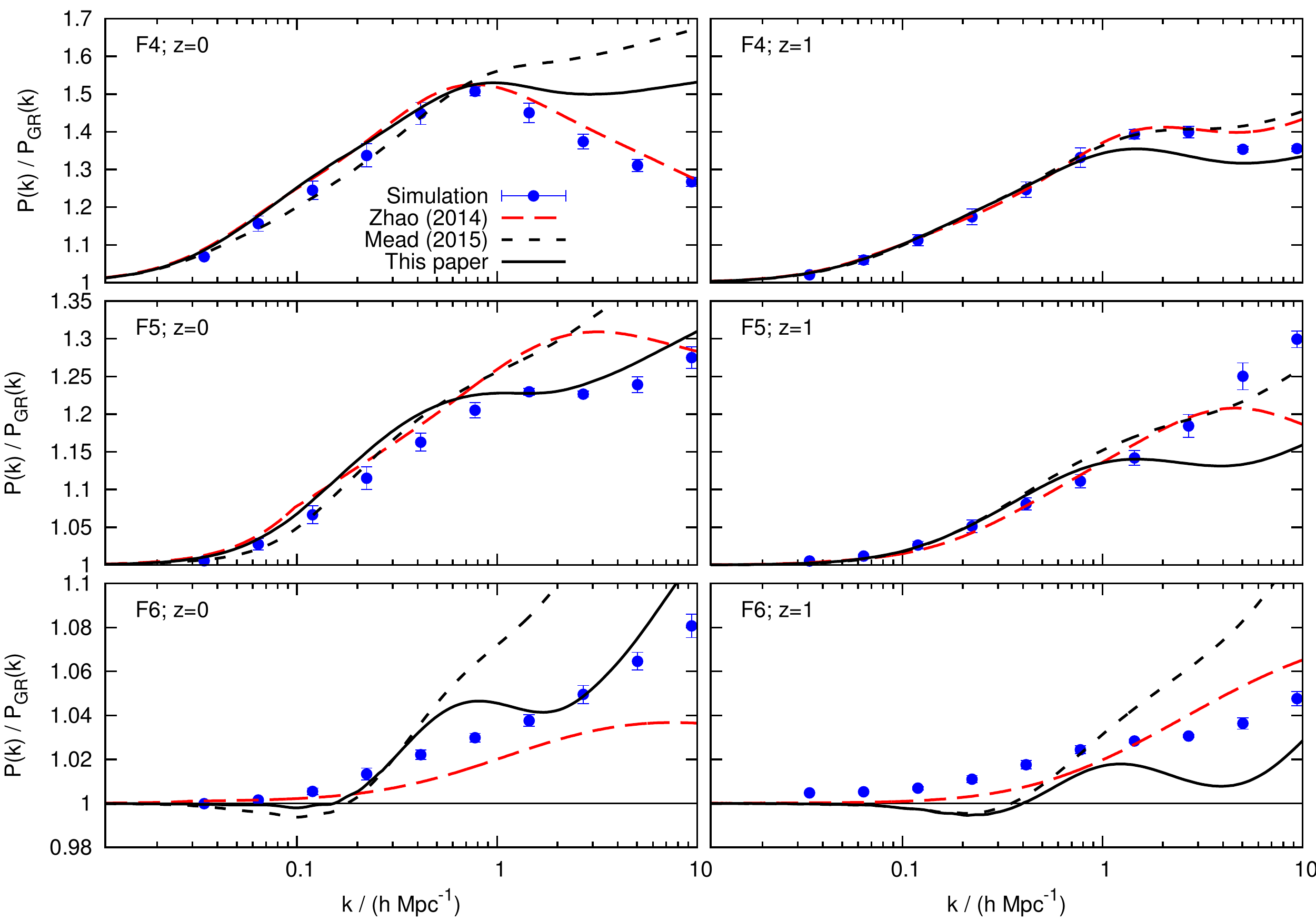}
\end{center}
\caption{A comparison of the ratios of power spectra for $f(R)$ models compared to an equivalent \LCDM model for $|f_{R0}|=10^{-4}$ (top), $10^{-5}$ (middle) and $10^{-6}$ (bottom) at $z=0$ (left-hand column) and $1$ (right-hand column). We show power from the simulations of \citeauthor{Li2013} (\citeyear{Li2013}; blue points) together that from two versions of the halo model; that of \citeauthor{Mead2015b} (\citeyear{Mead2015b}, short-dashed; black) and the update advocated in this paper (solid; black). We also show the \mghalofit model of \citeauthor{Zhao2014} (\citeyear{Zhao2014}; long-dashed; red) that was fitted to the same simulation data as shown (amongst others) and provides a better fit.}
\label{fig:fR}
\end{figure*}

%Simulations
We implement $f(R)$ models into our halo model using the modified scale-dependent growth function (solutions to equation~\ref{eq:perturbations}) and we use this to calculate $\sigma$ for the mass function. In Fig.~\ref{fig:fR}, we compare our results for the power spectrum response to the simulations presented in \cite{Li2013}, which are detailed in Table~\ref{tab:fR}. Although we can see that the \cite{Mead2015b} results are reasonable for $k<1\iMpc$, they fail to match the subtleties of the response as a function of $f_{R0}$. We also show the \halofit based extended fitting formula of \cite{Zhao2014}, which was tuned to these same simulations and fits the data reasonably well, although we note that the fitting function contains 39 free parameters in addition to the 38 from \cite{Takahashi2012}.

%Screening mechanism in spherical collapse
To improve halo-model predictions we use a simple model for chameleon screening that derives from spherical-collapse arguments. If $\Om=1$ and we simply take $G\to G_\mathrm{eff}$ (a constant) for perturbations, then we find good fits to spherical model results\footnote{The fact that the powers in equation~(\ref{eq:spherical_MG}) are the same as the multiples in equation~(\ref{eq:spherical_nu}) is not coincidental. In the limit in which we work, an unclustered massive neutrino background has an identical effect to a drop in $G$.} of
\begin{equation}
\eqalign{
\dc&\simeq 1.686\times (G_\mathrm{eff}/G)^{0.041}\ ,\cr
\Dv&\simeq 178\times (G_\mathrm{eff}/G)^{-0.763}\ .
}
\label{eq:spherical_MG}
\end{equation}
For $G_\mathrm{eff}/G=4/3$ these results are equivalent to those in \cite{Schmidt2009a} for the deviation from a \LCDM model. A simple model for screening in haloes is as follows: a region of the Universe can be considered screened when 
\begin{equation}
\bar{f}_{R}(a)\simeq \frac{2}{3} \Psi_\mathrm{N}\ ,
\end{equation}
as the $f_R$ field is forced into the minimum of the effective potential when the local gravitational potential is of the order of the background $f_R$ value \citep{Schmidt2010b}. This fact was used in \citeauthor{Mead2015a} (\citeyear{Mead2015a}; fig. 1) to develop a simple model for $G_\mathrm{eff}/G$ as a function of halo mass for NFW haloes, by relating the gravitational potential of a halo to the $f_{R}$ value. This toy calculation agrees well with results of full numerical calculations of screening in idealised symmetric haloes and \nbody simulations (\eg see fig. 3 of \citealt{Schmidt2010b}).

%Sigmoid
We then use a sigmoid approximation for $G_\mathrm{eff}/G$ as a function of halo mass:
\begin{equation}
%S(A,B,x_0,x_\mathrm{s},x)=A+\frac{1}{2}(B-A)\left[1+\tanh\left(\frac{x-x_0}{x_\mathrm{s}}\right)\right]
S(x-x_0,\Delta,A,B)=A+\frac{1}{2}(B-A)\left[1+\tanh\left(\frac{x-x_0}{\Delta}\right)\right]
\label{eq:sigmoid}
\end{equation}
which transitions between $A$ for $x\ll x_0$ to $B$ for $x\gg x_0$ with transition width governed by $\Delta$. This allows $G_\mathrm{eff}/G$ to transition from unscreened $4/3$ enhancement at low halo masses to screened at high mass:
\begin{equation}
%\frac{G_\mathrm{eff}}{G}=\frac{4}{3}-\frac{1}{3}\left\{\frac{1}{2}\left[\tanh\left(\frac{\log_{10}M-\log_{10}M_\mathrm{s}}{\log_{10}M_\mathrm{t}}\right)-1\right]\right\}\ .
\frac{G_\mathrm{eff}}{G}=S(\log_{10}M/M_\mathrm{s},\Delta,\frac{4}{3},1)\ ,
\label{eq:geff_sigmoid}
\end{equation}
with $M_\mathrm{s}$ a screening mass and a comparison of this function with the model in \cite{Mead2015a} leads us to fix the transition via $\Delta=0.38$ and
\begin{equation}
%M_\mathrm{s}=\sform{1.22}{22}\,\Om^{-1/2}(a)(-f_{R0})^{3/2}\,\Msun
\log_{10}\left(\frac{M_\mathrm{s}}{\Msun}\right)=14.6+\frac{3}{2}(5+\log_{10}|f_R(a)|)-\frac{1}{2}\log_{10}\Om \ .
\label{eq:screening_mass}
\end{equation}

%Fitting
We then fit our model $\dc$, $\Dv$ and $M_\mathrm{s}$ to the simulation response, taking the exponents in equation~(\ref{eq:spherical_MG}) and the amplitude of $M_\mathrm{s}$ in equation~(\ref{eq:screening_mass}) to be free parameters. Our best fitting model is shown as the solid line in Fig.~\ref{fig:fR} and is given by the relations
\begin{equation}
\eqalign{
\dc&\propto\mathrm{constant} \cr %[1+0.0123\log_{10}\Om(z)]\ , \cr
\Dv&\propto(G_\mathrm{eff}/G)^{-0.5}\ ,
\label{eq:spherical_MG_fit}
}
\end{equation}
and
\begin{equation}
\log_{10}\left(\frac{M_\mathrm{s}}{\Msun}\right)=18.9+\frac{3}{2}(5+\log_{10}|f_R(a)|)-\frac{1}{2}\log_{10}\Om \ .
\label{eq:screening_mass_fit}
\end{equation}
Note that fitting the halo model to simulation data prefers $\dc$ to have no dependence on $G_\mathrm{eff}$, but that the dependence of $\Dv$ is in the same sense and of a similar magnitude to the spherical model prediction. Also note that the leading factor in the fitted $M_\mathrm{s}$ expression means that our toy model prefers all haloes to be unscreened at $z=0$; although we caution the reader against taking this too literally. This improves accuracy to the $2$ per cent level for $k<1\iMpc$ and $10\%$ level for $k<10\iMpc$, apart from for the $z=0$ F4 model. The halo model is still lacking in accuracy compared to \mghalofit, but there is some indication that the accuracy of \mghalofit may degrade when away from the cosmological models on which it was trained \citep{Tessore2015} and we suggest that using a halo model approach may be preferable in general. Given the current statistical power of weak lensing, our level of accuracy is sufficient to test $f(R)$ models. It may be possible to improve our accuracy using more accurate prescriptions for spherical model parameters. Although previous results in this paper \citep[and others, \eg][]{Schmidt2010b,Lombriser2014a} should caution us against taking spherical model calculations as being relevant to power spectrum prediction (\ie one should \emph{not} expect a perfect calculation of $\dc$ or $\Dv$ to translate into perfect mass function or halo model power predictions).

%Spherical models in chameleon
Instead of equation (\ref{eq:spherical_MG}), the effect of the chameleon screening mechanism on the collapse density can be described by adopting a thin-shell approximation in the modified evolution equation of a spherical top-hat overdensity \citep{Brax2010,Li2012b}. This approach was used to compute $\dc$ for $f(R)$ gravity in \cite{Lombriser2013a}. Alternatively, \cite{Borisov2012} and \cite{Kopp2013} compute $\dc$ by considering an isolated, initial over-density profile set by peaks theory and its isotropic evolution according to equations~(\ref{eq:quasistatic_fr}) and (\ref{eq:psi}).

%Modelling power in chameleon models
A halo-model power spectrum based on the collapse density from the thin-shell approximation was used in \cite{Lombriser2014a} but this over predicts the modification in the power spectrum, although a correction term suppressing the enhancements in the two-halo term at quasilinear scales can be adopted to improve the description. Other approaches to modelling the power spectrum in $f(R)$ interpolate between the modified and screened Newtonian regimes. The interpolation can directly be modelled in the power spectrum as has been proposed by \cite{Hu2007b}. This transition was described through perturbation theory in \cite{Koyama2009} and an extension of it was fitted to \nbody simulations in \cite{Zhao2010}. An alternative interpolation was introduced by \cite{Li-Yin2011} who model the chameleon transition in $\sigma$ (equation \ref{eq:variance}) and fit the resulting halo mass function to \nbody simulations. They then adopt the halo model to describe the power spectrum with an interpolation between the two- and one-halo terms following the \halofit approach. \cite{Brax2013} used a combination of one-loop perturbations with a one-halo term to describe the power spectrum and \cite{Achitouv2015} incorporate the mass function model of \cite{Kopp2013} into halo-model predictions and achieve $20\%$ level matches to simulations for $k<3\iMpc$. A comparison between some of these different approaches can be found in \citeauthor{Lombriser2014b} (\citeyear{Lombriser2014b}; figs.~4 and 5). While some of these methods provide a better description of the $f(R)$ power spectra than the tuned halo model in this work, when comparing theoretical predictions to observations we are particularly interested in methods that efficiently compute the non-linear power spectrum and allow us to sample the entire available parameter space. Our halo model does not require spherical collapse or higher-order perturbation theory calculations and can therefore provide fast predictions. Due to its physical motivation, it may also be more easily generalized to other chameleon models besides the $f(R)$ modifications covered by \mghalofit.
%End of Lucas's bit

\subsection{Vainshtein screening}
\label{sec:Vainshtein}

%Introduction
The \cite{Vainshtein1972} screening mechanism is different from the chameleon, in that the screening depends primarily on local density, rather than halo mass. In this paper we work with a toy \citeauthor*{Dvali2000} (DGP; \citeyear{Dvali2000}) model that exhibits this screening. We work with the normal branch of the model, which does not self accelerate, but we impose an exact \LCDM background expansion on the model artificially. Once again, this means that the modification to gravity is not responsible for the acceleration, but allows us to study the MG in isolation from background effects. In practice, this model could be realized within the normal branch of DGP by adding a contrived, homogeneous DE to exactly cancel the non-standard background expansion that would otherwise be present. Throughout this section we use units such that $c=1$.
%Mention link to Galileon model

%Basic equations and 
In the quasi-static, sub-horizon limit the $\Psi$ and $\Phi$ potentials feel contributions from a new scalar field $\phi$:
\begin{equation}
\frac{1}{a^2}\nabla^2 \Psi=4\pi G\bar\rho\delta+\frac{1}{2a^2}\nabla^2 \phi\ ,
\label{eq:dgp_psi}
\end{equation}
\begin{equation}
\frac{1}{a^2}\nabla^2 \Phi=4\pi G\bar\rho\delta-\frac{1}{2a^2}\nabla^2 \phi\ .
\label{eq:dgp_phi}
\end{equation}
Once again, lensing is not directly affected by the modification because $\Psi+\Phi$ is unchanged from the standard. However, non-relativistic particles (accelerated by $\Psi$) feel $\phi$, the equation of motion of which is given by
\begin{equation}
\frac{1}{a^2}\nabla^2\phi+\frac{r_\mathrm{c}^2}{3\beta(a) a^4}\left[(\nabla^2\phi)^2-(\nabla_i\nabla_j\phi)(\nabla^i\nabla^j\phi)\right]=\frac{8\pi G}{3\beta(a)}\bar\rho\delta\ ,
\label{eq:dgp_scalar}
\end{equation}
\citep[\eg][]{Koyama2007} where we have employed summation convention over spatial coordinates. $r_\mathrm{c}$ is a free parameter in the model and is known as the `crossing radius'; the larger $r_\mathrm{c}$ the less strong the modification. The function $\beta(a)$ is given by
\begin{equation}
\beta(a)=1+\frac{4}{3}Hr_\mathrm{c}\left(1+\frac{A}{2H^2}\right)\ ,
\label{eq:dgp_beta}
\end{equation}
where $H$ is the Hubble function and $A\equiv\ddot a/a$. Note that $\beta>0$ always and larger $\beta$ implies less modification. Since $\beta(a)$ was larger in the past, the modification is only felt at late times. For standard \LCDM backgrounds the value of $\beta$ saturates as the Universe reaches the de Sitter phase.

%Linear theory
If we linearize for small values of $\phi$, the $\Psi$ Poisson equation reads
\begin{equation}
\frac{1}{a^2}\nabla^2 \Psi=4\pi G\left[1+\frac{1}{3\beta(a)}\right]\bar\rho\delta\ ,
\label{eq:dgp_linear}
\end{equation}
and we see the linear gravity modification is time dependent, but scale independent.

%Vainshtein screening
Insight into the non-linear action of the Vainshtein mechanism can be gained from either simulations \citep[\eg][]{Schmidt2010b,Falck2015} or spherical calculations \citep[\eg][]{Schmidt2010a}. These reveal that in regions of sufficiently high density $\nabla\phi\ll\nabla\Psi$ and screening has occurred. For spherically symmetric density profiles, an exact solution exists for the radial derivative of the field (\ie the force modification):
\begin{equation}
\derivative{\phi}{r}=\frac{3\beta(a) r}{4 r_\mathrm{c}^2}\left[\sqrt{1+\left(\frac{r_*(r)}{r}\right)^3}-1\right]\ ,
\end{equation}
where the Vainshtein radius has been defined
\begin{equation}
r_*(r)=\left(\frac{16GM(r)r_\mathrm{c}^2}{9\beta^2(a)}\right)^{1/3}\ ,
\end{equation}
and $M(r)$ is the mass enhancement, relative to the background, enclosed at radius $r$. For $r\ll r_*$, the modification to gravity is negligible. Note that `the' Vainshtein radius is actually a function of $r$. Outside the halo $M(r)\rightarrow M$ and the Vainshtein radius scales $\propto M^{1/3}$, which is exactly the scaling of the virial radius, and this means that the Vainshtein mechanism is roughly independent of halo mass; although it does depend on the details of the halo profile to some extent. Also note that these equations conspire to make the efficiency of screening at $r$ a function of the enclosed, average density at $r$. For cosmologically interesting values of $r_\mathrm{c}$ the force modifications are only felt further out than the halo virial radius, allowing these models to pass Solar system and Galactic gravity tests.

%Environment dependence
Note that the non-linear modification depends on density morphology through the structure of the derivatives in the quadratic $\phi$ term in equation~(\ref{eq:dgp_scalar}). For example, in situations of planar symmetry the quadratic $\phi$ term vanishes and the screening mechanism is annihilated simply due to the symmetry. Therefore, different environments (\eg clusters, filaments, sheets, voids) feel different levels of modified forces. However, \cite{Schmidt2010b} and \cite{Falck2015} show that although morphology screening is active for general DM particles, it is negligible for particles inside haloes because it is the spherical nature of the halo that trumps over the environment of the halo: thus all haloes are screened independently of environment.

\begin{table}
\centering
\caption{Simulations of standard gravity (GR) and DGP models analysed in this paper. The cosmological parameters are $h = 0.703$, $\Om =0.271$, $\Ob = 0.045$, $\Ov = 1-\Om$ and $n_\mathrm{s} = 0.966$. Structure formation takes place in a cube of side $250\Mpc$ from exactly the same initial conditions. Linear growth is more enhanced in models with smaller $r_\mathrm{c}$ values.}
\begin{tabular}{c c c c}
\hline
Simulation & $H_0 r_\mathrm{c}$ & $r_\mathrm{c}/(\Mpc)$ & $\sigma_8$ \\ [0.5ex] 
\hline
GR & $\infty$ & $\infty$ & 0.800  \\
DGP1 & 0.5 & 1,500 & 0.896  \\
DGP2 & 1.2 & 3,600 & 0.849  \\
DGP3 & 5.6 & 16,800 & 0.812 \\
\hline
\end{tabular}
\label{tab:DGP}
\end{table}

\begin{figure*}
\begin{center}
\includegraphics[angle=270,width=18cm]{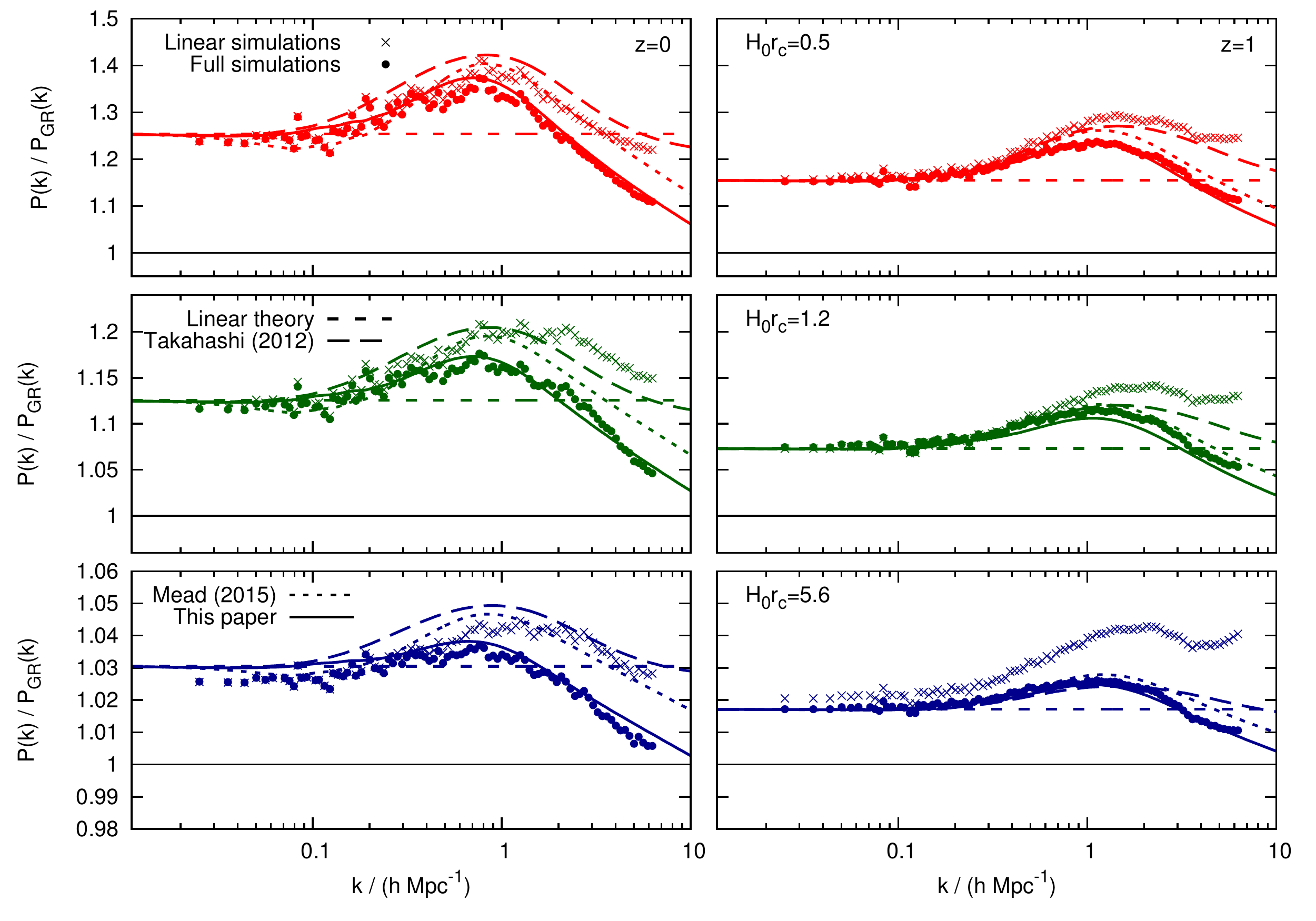}
\end{center}
\caption{A comparison of the power spectrum response for DGP models with the same expansion history as \LCDM but a time-varying gravitational constant on large scales, but which is screened by the Vainshtein mechanism within haloes. We show power from full simulations (filled circles) and linear simulations (crosses) for three models: $H_0 r_\mathrm{c}=0.25$ (top; red) $1.2$ (middle; green) and $5.6$ (bottom; blue) at $z=0$ (left column) and $1$ (right column). Also shown are the linear enhancement (medium-dashed) predictions from \halofit of \citeauthor{Takahashi2012} (\citeyear{Takahashi2012}; long-dashed), predictions from \citeauthor{Mead2015b} (\citeyear{Mead2015b}; short-dashed) as well as the fitted halo-model prediction in this work (solid).}
\label{fig:DGP}
\end{figure*}

%Simulations
We utilize simulations for three different values of $r_\mathrm{c}$ as well as for a \LCDM model with matched initial conditions, these are listed in Table \ref{tab:DGP} and were run using the simulation code \textsc{isis} \citep{Llinares2014}. The linear gravity enhancement means that large-scale structure is more developed in the Vainshtein models compared to \LCDM. We compute the power spectrum for each simulation at $z=0$ and show the response in Fig.~\ref{fig:DGP}. The short-dashed lines show the expected, constant enhancement from linear perturbation theory, which the simulations follow with their largest modes ($k\leq 0.1\iMpc$). The long-dashed lines show the power spectrum prediction of \halofit, which we see matches the standard bump enhancement around $k\simeq 0.2\iMpc$ for each model. However, we see that this enhancement is suppressed in the simulations relative to the \halofit expectation, and is almost absent in the $H_0 r_\mathrm{c}=5.6$ model. The same is true of predictions from the halo model presented in \cite{Mead2015b}. Clearly this defect in modelling arises because we have not included the Vainshtein mechanism at all, which ensures that gravitational forces on small scales return to standard. To ameliorate this we once again turn to the spherical model.

%Spherical model
The spherical model in DGP gravity was investigated in \cite{Schmidt2010a} where it was found that improved mass function, halo bias and halo-model power spectra predictions were produced when incorporating spherical model predictions for $\dc$ and $\Dv$. However, the results are not perfect for the power spectrum, and not accurate enough to be used in an analysis. We employ the same strategy in this work, but permit ourselves the freedom of altering the amplitude of the spherical model $\dc$ and $\Dv$ relations to best match the simulated power spectrum data. We also ignore the Vainshtein mechanism in our spherical model calculation, because \cite{Schmidt2010b} show that it makes very little difference to the values of $\dc$ and $\Dv$. Our hope is that allowing ourselves the freedom of altering the amplitude of the deviation in the spherical model calculations will account for the Vainshtein mechanism and produces accurate results. We find that DGP Einstein-de-Sitter spherical model results (see Appendix~\ref{app:spherical_model}) can be well fitted by
\begin{equation}
\eqalign{
%\dc&\simeq 1.686\times[1-0.003a(H_0 r_\mathrm{c})^{-0.7}]\ , \cr
\dc&\simeq 1.686\times S(\log_{10}[2.33a/(H_0r_\mathrm{c})^{0.65}],0.4,1.,0.9973)\ , \cr
\Dv&\simeq 178\times[1-0.08a(H_0 r_\mathrm{c})^{-0.7}]\ ,
}
\label{eq:spherical_nDGP}
\end{equation}
where $S$ is the sigmoid function defined in equation (\ref{eq:sigmoid}). We then fit the halo model power spectrum to simulation data letting the amplitude of the $\dc$ and $\Dv$ relations be free parameters (the $0.9973$ in the $\dc$ equation and $-0.08$ in that of $\Dv$). We find best fitting amplitudes for $\dc$ of $1.008$ and $-0.0388$ for $\Dv$. Our final power spectra are presented in Fig. \ref{fig:DGP} (solid lines), where it can be seen that we achieve few per cent level matches to all three models at $z=0$ and $1$. Once again, the trend in $\dc$ preferred by the halo model fitted to the simulation data is opposite to that of spherical model calculations. \cite{Schmidt2010b} show improved response from the halo model when using the spherical model to calculate both $\dc$ and $\Dv$, but the resulting response is by no means accurate.
%This seems strange since Schmidt gets better predictions using dc from the spherical model (different halo model implementations?) I think the difference (improvement) is mainly drive by Dv, which is also what we see.

%Galileon simulation work
The spherical model in Galileon models (which are also Vainshtein screened) has been investigated by \cite{Barreira2013} and the Galileon halo model by \cite{Barreira2014}. For the mass function they find a need to recalibrate the parameters of \cite{Sheth1999} to obtain good results to the mass function in their simulations. The eventual halo model power spectra presented are only accurate at the $20\%$ level even when using relations (\eg mass function, $c(M)$) calibrated from the exact simulations they compare to for the halo model power spectrum. Again, this indicates that the halo model in its unaltered form is \emph{not} an accurate tool for predicting the non-linear spectrum, we suppose primarily because of features in the non-linear density field that are missing from the model.

\section{Degeneracy}
\label{sec:degeneracy}

\begin{figure}
\begin{center}
\includegraphics[angle=270,width=8.5cm]{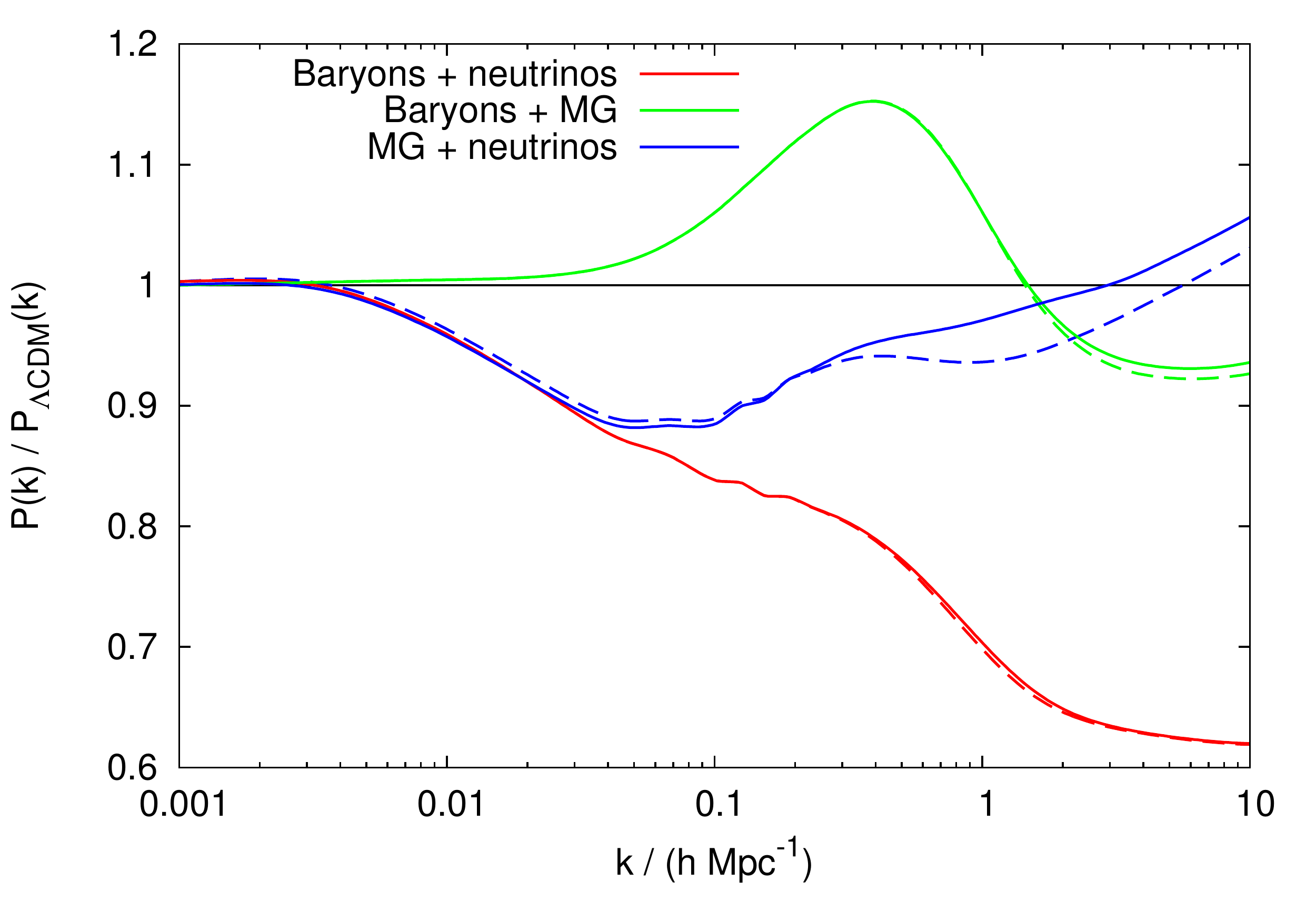}
\end{center}
\caption{The ratio of power compared to a fiducial \LCDM model for cosmological models with combinations of massive neutrinos ($\sum_\nu m_\nu=0.3\eV$), baryonic feedback (an AGN model) and MG ($f_{R0}=-10^{-5}$ and $n=1$). We show combinations of two of the three effects where the ratio is computed both by assuming the effects act independently (dashed line) and via the full halo model calculation (solid line) where the effects are treated in tandem. Differences between each pair of dashed and solid lines indicate the extent to which the effects can be treated independently. We see that this is a good approximation in the case of the combination of baryonic feedback with either MG or massive neutrinos. However, the approximation degrades when one considers MG and massive neutrinos together; the non-linear neutrino suppression effect is weaker in $f(R)$ compared to standard gravity.}
\label{fig:degeneracy}
\end{figure}

%Method
The modelling developed in the previous section allows us to investigate degeneracies in the matter power spectrum between extensions to the standard cosmological paradigm. Simulations that investigate these degeneracies are rare due to their complexity and expense, but notable exceptions are presented in \cite{Puchwein2013}, \cite{Arnold2014}, \cite{Baldi2014} and \cite{Arnold2015}. However, we may use the halo model to compute the power spectrum for combinations of extensions, and we may also include the effects of baryons using the method presented in \cite{Mead2015b}. Our results are presented in Fig.~\ref{fig:degeneracy} where we show power spectra for combinations of massive neutrinos ($M_\nu=0.3\eV$ from the modelling in Section~\ref{sec:massive-nu}), baryonic feedback (the AGN halo model from \citealt{Mead2015b}; obtained by altering parameters within the halo model that govern halo internal structure \new{so as to match power spectra from the OWLS simulations of \citealt{Schaye2010,vanDaalen2011}}) and MG (an $f(R)$ model with $f_{R0}=-10^{-5}$ and $n=1$ from Section~\ref{sec:chameleon}). For each pair of extensions, the solid line show the result of a full halo-model calculation, with the combined effects, whereas the dashed line shows the result of using the halo model, but assuming that the extensions to the standard paradigm can be treated independently. The difference between each pair of lines for the different combinations indicate how good an approximation it is to treat the effects independently. Recent work \citep[\eg][]{Harnois-Deraps2015a,Harnois-Deraps2015c} looking at cosmological constraints from all three effects in tandem has been forced to assume they act independently, due to the lack of theoretical models, but with the work presented in this paper we are now in a position to model these effects simultaneously. The linear spectrum for the `$f(R)$ and massive neutrino' case is calculated using \mgcamb \citep{Hojjati2011,MGCAMB}, but we find that the approximation of treating the two effects separately and then combining them is almost perfect for the linear spectra (within a per cent for $k<10\iMpc$). For this model, we compute $\dc$ and $\Dv$ by multiplying the deviations from equations~(\ref{eq:spherical_nu_fit}) and (\ref{eq:spherical_MG_fit}). This is justified because the origin of the deviation is the same in each case being due to either an increased or a decreased effective $G$, and is small in any case. Since the spherical model calculation is non-linear, it may seem preferable to repeat it for each combined model. However, we do not do this since we found that the spherical model calculations did not provide accurate halo-model power spectra. 

%Baryons
From Fig.~\ref{fig:degeneracy}, we see that the halo model predicts that baryonic feedback should act quite independently from either massive neutrinos or MG. Physically, this can be understood as feedback affecting halo internal structure only, but not altering the mass function or large-scale clustering to the same extent \citep[\eg][]{vanDaalen2014}. Thus the power spectrum response predicted by considering feedback in isolation is a fairly good approximation to the response in more general models. This is in qualitative agreement with results presented by \cite{Puchwein2013} and may be good news considering the relative uncertainty and expense of simulations that contain hydrodynamics and baryonic feedback; it may be sufficient to run detailed hydrodynamic simulations with feedback for standard \LCDM models and then to translate these effects into extensions of the standard model. In contrast, from Fig.~\ref{fig:degeneracy} we see much larger differences when considering massive neutrinos and MG in tandem. \cite{Baldi2014} investigate \nbody simulations that include both massive neutrinos and \cite{Hu2007b} $f(R)$ gravity (their fig. 4) and come to a similar conclusion. Our results are in qualitative agreement with those from the full simulations, in that we see that the suppression of power caused by massive neutrinos is less in the $f(R)$ models compared to standard gravity. This is despite the multiplication of ratios being a very good approximation to create the linear spectrum for these models. Physically this may be because both mechanisms effect the linear growth of perturbations, and this translates into the non-linear density field in a non-linear way. For example, both mechanisms affect the mass function, but not in such a way that the induced power spectrum change is well approximated by considering each separately. The level of this non-linear coupling is not sufficient to have affected the results of \cite{Harnois-Deraps2015a,Harnois-Deraps2015c} and will probably not need to be modelled for near-term lensing surveys. However, lensing surveys that will be active in the next decade have stringent, per cent level accuracy requirements on the theoretical modelling of the power spectrum in order to provide forecasted constraints. Our results indicate that it will not be sufficient to consider these extensions to the cosmological paradigm in isolation.

\section{Summary and discussion}
\label{sec:summary}

%Summary of results
We have demonstrated that it is possible to use the halo model to provide accurate matter power spectrum predictions for a variety of extensions to the standard cosmological paradigm. Starting from the accurate halo model of \cite{Mead2015b} we investigated both minimally and non-minimally coupled DE models, massive neutrinos and MG forces. In each case we compared predictions from the halo model to simulations of each extension at the level of the power spectrum response; a measure of deviation from a fiducial model.

%Dark energy
We demonstrated that models of DE parameterized via $w(a)=w_0+(1-a)w_a$ could be matched at the few per cent level for $k<10\iMpc$ with no modifications to the \cite{Mead2015b} halo model and we noted that our predictions are more accurate than those the \cite{Takahashi2012} version of \halofit. This suggests that our halo model would produce reasonable answers for other quintessence models, simply by using the correct linear growth function and power spectrum. We were able to produce few per cent level accuracy for coupled quintessence models for $k<1\iMpc$, but to extend to smaller scales we were forced to re-fit the concentration--mass relation that enters into the halo model calculation to power spectrum data for each coupled model. While this is unsatisfactory, we take comfort in the fact that the required relation at least reflects that measured in these simulations. If some way were developed to predict this relation from first principles then we feel confident that an accurate halo-model prediction scheme could be developed. 

%Massive neutrinos
For massive neutrinos, we adapted our halo model calculation so that quantities relevant to structure formation were computed using only the cold, clustering component of the matter only (\ie ignoring the hot neutrino component). Results from this approach were satisfactory on small scales, but we observed that the \cite{Mead2015b} halo model under predicted the power spectrum suppression caused by massive neutrinos around quasi-linear scales ($k\simeq 0.3\iMpc$). To remedy matters we re-parameterized the Zel'dovich damping term in the \cite{Mead2015b} halo model such that results presented in that paper were unaffected (even improved), but so that we matched the quasi-linear damping in massive neutrino cosmologies. To further improve predictions at smaller scales we turned to the spherical model, utilizing the simplifying assumption that neutrinos are completely unclustered in the background, which corresponds exactly to a reduced gravitational constant for clustering. While the exact form of the predictions for the linear-collapse threshold ($\dc$) and virialized over-density ($\Dv$) did not provide improved power spectrum predictions, we were able to use the $f_\nu$ dependence from the spherical model results as the basis for a fitting function that \emph{does} provide improvement. However, the trend in $\dc$ preferred by our fitting to power spectrum data is opposite to that in spherical-model calculations, whereas the trend in $\Dv$ is of the same sense and magnitude. It may be that our spherical model is too simplistic, and that if we were to include neutrino clustering and free-streaming \citep[\eg][]{Ichiki2012,LoVerde2014} our result for $\dc$ may be fundamentally different. Certainly a neutrino perturbation in addition to that of CDM will accelerate collapse, but results from \cite{Ichiki2012} indicate that this accelerated collapse is not sufficient to reverse the trend in $\dc$. Regardless, our final accuracy in comparison to simulations with massive neutrinos is at the few per cent level for $k<10\iMpc$ and offer a slight improvement over the \cite{Bird2012} extension to \halofit.

%Chameleon
We investigated non-linear power spectrum predictions for chameleon models taking \cite{Hu2007b} $f(R)$ models as an example. Once again, we turned to the spherical model for a way to parameterize functional forms for $\dc$ and $\Dv$ that we could use to improve predictions. In chameleon models, screening is a function of halo mass, and we introduced a fitting formula for spherical model parameters that included a `screening mass' below which haloes feel a $4/3$ gravity enhancement and their spherical model parameters are adjusted accordingly. Even when fitting this three-parameter model to simulation data, we were only able to produce $10$ per cent level matches for $k<10\iMpc$. While the deviation in models with smaller $|f_{R0}|$ values is smaller, our model for the deviation is also a poorer fit to simulated data for these smaller values. Also, the simulation data prefer the absence of a change in $\dc$ for $f(R)$ models while the trend in $\Dv$ is of a similar magnitude and the same sense as the spherical model prediction. Our model for chameleon screening in the halo model is perhaps too simple, and in the future it may be fruitful to consider more complicated spherical models \citep[\eg][]{Li2012b,Kopp2013} or some more universal model for screening such as that proposed by \cite{Gronke2015} to see if these result in improved power spectrum predictions when using the same fitting approach we advise in this work. For example, it may be useful to incorporate environmental screening and the fact that some haloes transition between being screened and unscreened as the universe evolves.

%DGP
We also investigated the example Vainshtein screened MG model of \citeauthor*{Dvali2000} (DGP; \citeyear{Dvali2000}) and were able to provide few per cent level accuracy compared to simulation data with fitting formula for the spherical model parameters. Once again, the trend in $\dc$ required to fit simulation power spectrum data is opposite to that predicted by the spherical collapse model, whereas that for $\Dv$ is of a similar magnitude and the same sense. The relatively good performance of our halo model for Vainshtein screened models compared to chameleon may be because the screening mechanism itself is simpler than the chameleon, and all haloes are essentially screened for all times in the model.

%Delta-v
It is difficult to know how to interpret our fitting formula in terms of spherical model results: we have seen that to get accurate power spectra we usually require a trend in $\dc$ that is opposite to that predicted by the spherical model. This contradiction may arise because we are using a tuned version of the halo model, rather than a pure one. However, it has been shown by many other authors that including the `correct' spherical model predictions within a halo model power spectrum calculation often leads to only marginal improvements \citep[\eg][]{Schmidt2009a,Schmidt2010a,Barreira2014}.

It has been shown that the mass function is approximately universal for a set halo definition ($\Dv$ either via a friends-of-friends linking length or spherical over-density criterion) but some authors say that universality breaks down if different $\Dv$ (\eg from the spherical model) are used to define haloes in different cosmologies \citep[\eg][]{Jenkins2001,Tinker2008,Courtin2011} even though enhanced universality might be expected. \new{Although recently, \cite{Despali2016} have claimed that the mass function appears to be universal if the spherical model virial density is used to define haloes in a cosmology dependent way via the SO algorithm.}  However, the usual inconsistency between halo definition and spherical model prediction means we cannot make direct comparisons between haloes identified in simulations and the spherical model. Given this, it then seems strange (and lucky) that the mass function is so close to universal for a cosmology independent $\dc$ and halo definition. Indeed, using the `correct' value of $\dc$ in a mass function prediction such as that of \cite{Sheth1999} seems to provide only marginal gains at the high mass end \citep{Courtin2011} and certainly does not fix remaining non-universality in the mass function. Comparing spherical model $\Dv$ predictions to simulations is hard because a $\Dv$ criterion is used to define a halo in the first place. \new{Therefore, it is difficult for us to make recommendations about where improved modelling efforts should be focused, but we suggest that insight may be gained by investigations into the differences in non-linear structure between models that are similar at the linear level \cite[\eg][]{McDonald2006} because this allows one to isolate the cosmology dependence from effects that arise from power spectra of different shape or amplitude.}

%A comparison of the halo model power spectrum to that measured in simulations could be used to indirectly probe $\Dv$ and $\dc$, but only under the standard assumptions of the halo model (\ie that all matter is contained in spherical haloes), which certainly does not hold if one is interested in per cent level accuracy rather than a broad brush picture of non-linear structure formation.

%Spherical model
The spherical top-hat is incredibly useful as an analytically soluble non-linear structure formation model and the predictions $\Dv\simeq 200$ and $\dc\simeq1.686$ are held firmly in the mind of any large-scale structure cosmologist. The \cite{Press1974} model predicts universality in the mass function, and this model and its extensions work absurdly well for halo formation, the mass function and halo bias given the simplistic assumptions. However, universality primarily arises through $\sigma$ in $\nu=\dc/\sigma$ and what is less obvious is that if accurate calculations of $\dc$ are relevant for improving mass function predictions or halo-model power spectra, or that if the number $1.686$ should be used as a guideline. Indeed, most modern mass functions \citep[\eg][]{Sheth1999,Warren2006,Tinker2008} are parameterized only in terms of $\sigma$; $\dc$ is either not present of phased out by some fitting parameters. 

%It has been known for a long time that the mass function is only approximately universal \citep[\eg][]{Sheth1999} and it is clear that spherical model calculations of $\dc$ do not fully alleviate deviations from universality. The spherical model has also been investigated in massive neutrino models by \cite{Ichiki2012,LoVerde2014} and used in mass function and halo model calculations for MG by \cite{Schmidt2009a,Schmidt2010a,Lombriser2013a,Barreira2013,Barreira2014} amongst others. In some cases the trends predicted by the spherical model improve mass function predictions and in others not. %In some cases the halo-model power spectrum is improved and in others not and it is difficult to disentangle the effects of a change in $\dc$ from a change in $\sigma(R)$ because both are typically modified.

%Blab
So, are we to conclude that the details of spherical model calculations are irrelevant for the details of the mass function and halo-model power spectrum? The spherical model is simplistic, and clearly the minutiae of non-linear structure formation are far from the minutiae of a collapsing top-hat, and may even be unrelated. It may be that in order to progress a more complicated model needs to be devised. One example is the ellipsoidal collapse model, which has been shown to better reflect the mass function by \cite*{Sheth2001} and more recently to provide an accurate model of halo concentrations by \cite{Okoli2016}. Though even this model is simplistic when compared to the chaos of true non-linear collapse. Maybe we have reached the limits of what we can learn about non-linear structure formation from simplistic calculations and detailed understanding can only come from the detailed analysis of simulations?

%Degeneracy
Finally, we investigated non-linear degeneracies between the extensions we consider, with a particular emphasis on investigating the approximation that extensions beyond the standard cosmological paradigm can be considered independently. Our results indicate that is it a reasonably good assumption to consider baryonic feedback processes in isolation from either MG or massive neutrinos, probably because feedback affects the internal structure of haloes, while leaving their numbers unaffected. In contrast, we found that our halo model predicts a non-negligible coupling between $f(R)$ gravity and massive neutrinos, such that a full halo-model power spectrum is different from one where the effects were treated independently at the $5$ per cent level for $k>0.5\iMpc$. Relatively few \nbody simulations have been run that investigate combinations of beyond standard model physics, but we note that our halo-model results are in qualitative agreement with those in the joint $f(R)$-massive neutrino simulations of \cite{Baldi2014}. This may have profound implications for the modelling effort required to extract accurate cosmological parameter constraints from so-called Stage IV lensing surveys that will be active in the next decade.

An updated version of our publicly available halo-model power spectrum generation code (\meadfit) is available at \meadaddress. This version covers the $w(a)$ DE models discussed in this paper, but not massive neutrinos or MG. \new{A version of \meadfit that includes massive neutrinos is incorporated within \camb\footnote{http://camb.info/}, and can be found at \texttt{https://github.com/cmbant/CAMB} and will be released with the next public \camb release.}

\section*{Acknowledgements}

AJM acknowledges support from a CITA National Fellowship and, together with CH, support from the ERC under the EC FP7 grant number 240185. CH also acknowledges ERC support under grant number 647112. LL has been supported by the STFC Consolidated Grant for Astronomy and Astrophysics at the University of Edinburgh and from an Advanced Postdoc. Mobility Grant from the Swiss National Science Foundation (no. 161058). We thank: Peter Ballett and Alex Hall for useful discussions regarding neutrinos, Marco Baldi for making the \codecs data public, Elena Massara and Francisco Villaescusa Navarro for providing the massive neutrino power spectrum data, Shankar Agarwal for making \pkann public and Baojiu Li for providing the $f(R)$ power spectrum data and all for useful discussions.

\label{lastpage}

%All of this stuff is to make the bibliography look better
%Enclose in 'footnote size' to make it smaller font size
\footnotesize{
%JP added these two things to left align it properly
\setlength{\bibhang}{2.0em}
\setlength\labelwidth{0.0em}
\bibliographystyle{mnras}
\bibliography{./../meadbib}
}

\normalsize
\appendix

\section{Simple spherical model calculations}
\label{app:spherical_model}

In this appendix we give our method and results for the spherical model calculations that are used as the basis for fitting formula that are used in this paper.

In the spherical model we are interested in solving the full non-linear equations that govern the evolution of a spherical top-hat overdensity. Under the assumption that a spherical hat remains a spherical hat throughout its evolution, the equation of motion for the over-density is
\begin{equation}
\ddot\delta+2H\dot\delta-\frac{4}{3}\frac{\dot\delta^2}{1+\delta}=\frac{3}{2}\Om(a)H^2\delta(1+\delta)\ .
\label{eq:spherical_tophat}
\end{equation}
In an Einstein-de-Sitter universe, an initially overdense top hat will always collapse at some point in the future, but this is only true in \LCDM if the initial perturbation is sufficiently dense. The time of collapse is defined as the time at which $\delta\rightarrow\infty$. On linearizing equation~(\ref{eq:spherical_tophat}) we recover the standard equation for the evolution of an arbitrary perturbation configuration
\begin{equation}
\ddot\delta+2H\dot\delta=\frac{3}{2}\Om(a)H^2\delta\ .
\label{eq:linear_tophat}
\end{equation}
The linear collapse threshold, $\dc$, is defined as the value that the linear field has reached at the time that collapse occurs in the full non-linear model. In practice, this is calculated by simultaneously evolving equations~(\ref{eq:spherical_tophat}) and (\ref{eq:linear_tophat}) for the same initial condition: a small seed over-density evolving in the linear theory growing solution (note this is $\delta\propto a$ only if $G_\mathrm{eff}=1$). The virialized over-density is calculated assuming that the virial theorem applies at the time of collapse; essentially that there is the virial split of kinetic and potential energy. This determines the radius of the collapsed top-hat that may then be translated into $\Dv$. For models with $\Lambda$ (or any DE) one should also consider the contribution of $\Lambda$ to halo support, but our calculations only consider $\Om=1$ and $\Ov=0$ models.

\begin{figure}
\begin{center}
\includegraphics[angle=270,width=8.5cm]{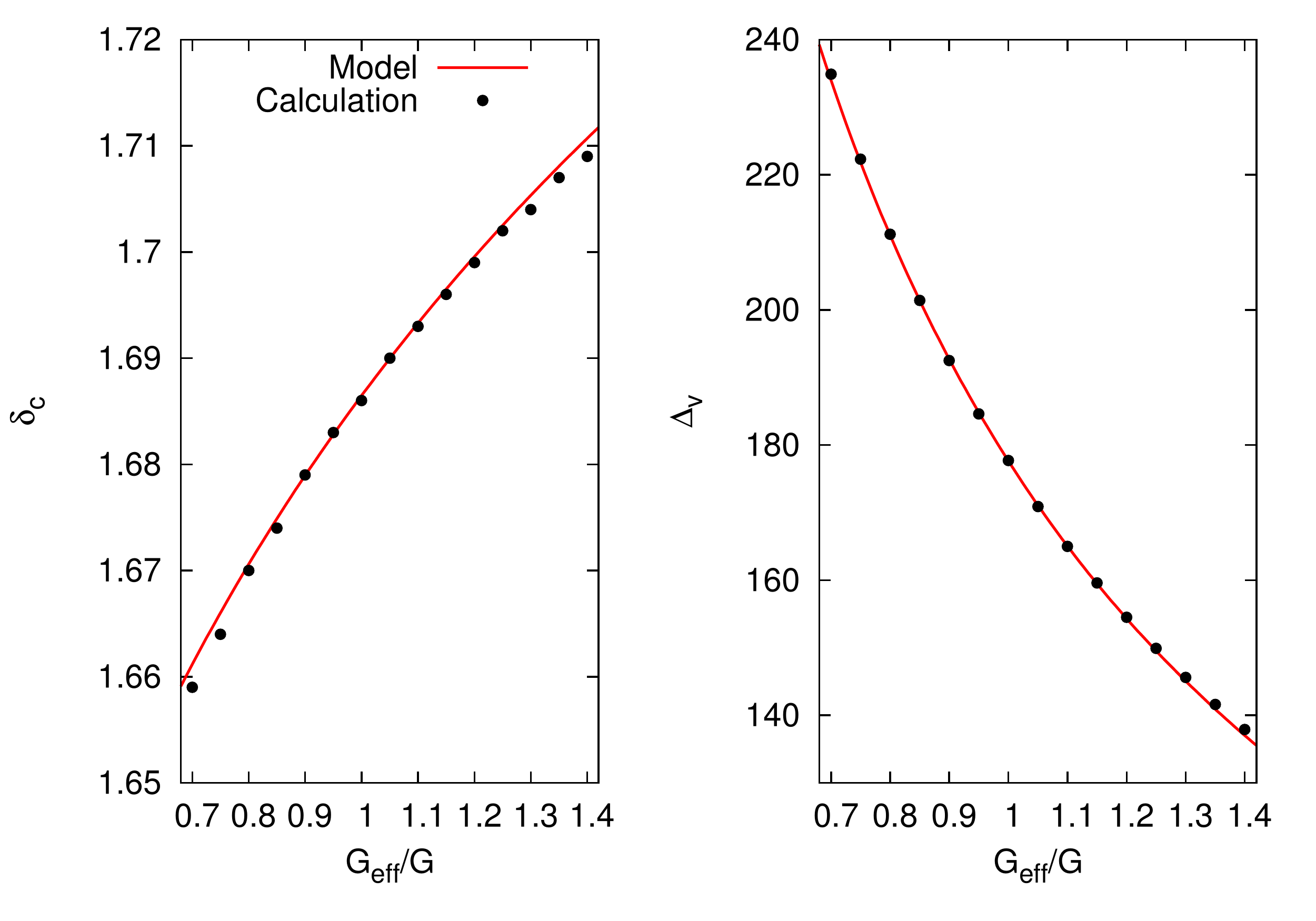}
\end{center}
\caption{The results of a spherical model calculation (black dots) and a fit to those results (red lines) for models where gravity is enhanced or suppressed for perturbations by a constant factor for all scales and at all times. The background expansion is standard, with $\Om=1$ and $\Ov=0$. At this level the effect of $G_\mathrm{eff}$ can be mapped directly into having a smoothly distributed matter component via $G_\mathrm{eff}=G(1-f_\nu)$ and an increase in $f_\nu$ is exactly equivalent to a lower gravitational constant for perturbations.}
\label{fig:MG_spherical}
\end{figure}

In our simplified calculations for massive neutrinos, the right-hand sides of equations~(\ref{eq:spherical_tophat}) and (\ref{eq:linear_tophat}) are multiplied by $1-f_\nu$ to reflect the fact that the fraction of mass in neutrinos does not contribute to gravitational clustering on small scales. For a constant, enhanced gravitational constant the right-hand sides are multiplied by $G_\mathrm{eff}/G$, which arises because the right-hand sides derive from the gravitational Poisson equation $\nabla^2\Phi=4\pi G\bar\rho\delta$. Note that this means increasing $f_\nu$ is exactly equivalent to a drop in $G$. The spherical model parameters that result from a change in the gravitational constant for perturbations are shown in Fig.~\ref{fig:MG_spherical}. The fitting functions are power laws
\begin{equation}
\eqalign{
\dc&\simeq 1.686\times (G_\mathrm{eff}/G)^{0.041}\ ,\cr
\Dv&\simeq 178\times (G_\mathrm{eff}/G)^{-0.763}\ .
}
\label{eq:spherical_MG_appendix}
\end{equation}

\begin{figure}
\begin{center}
\includegraphics[angle=270,width=8.5cm]{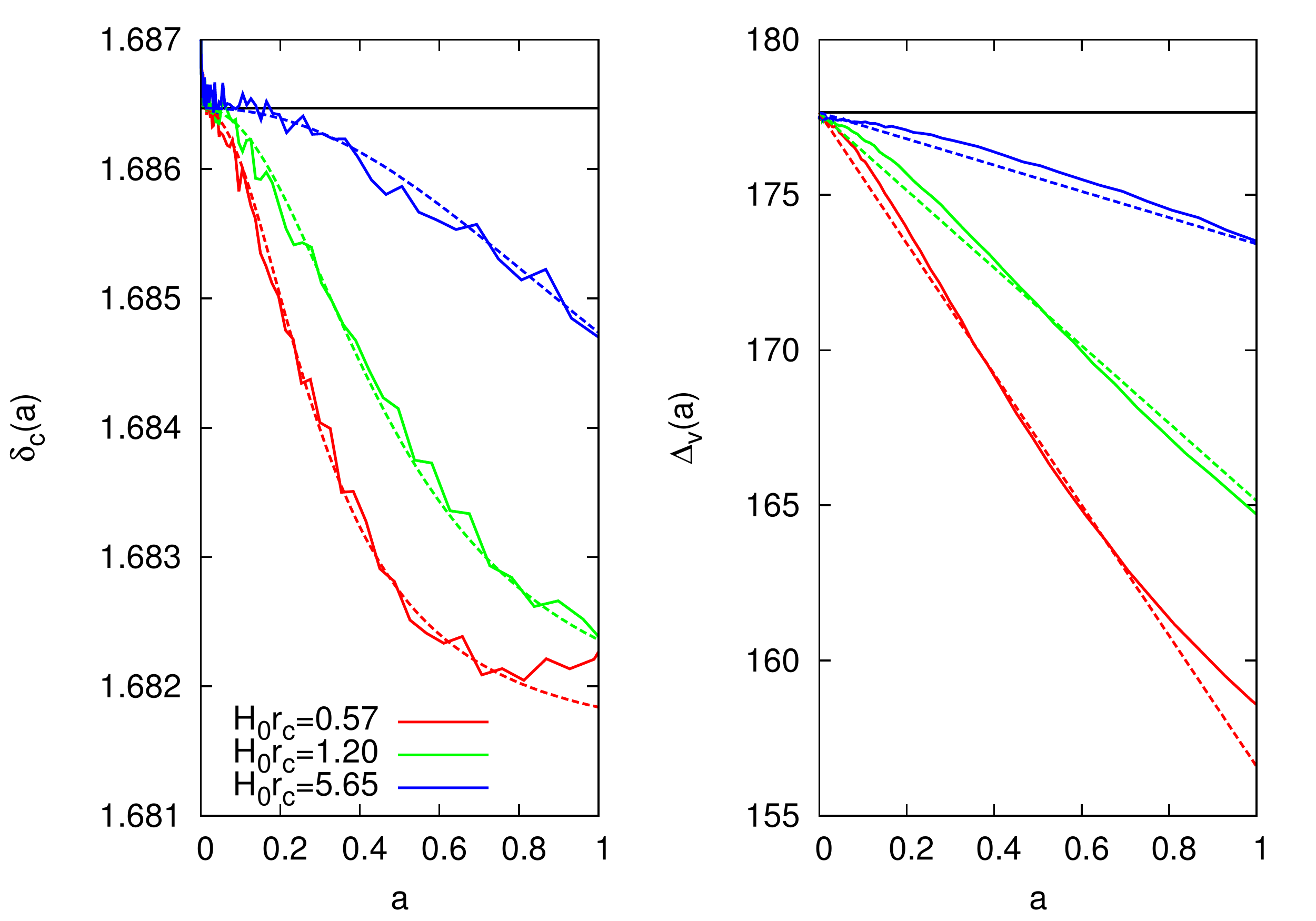}
\end{center}
\caption{Results from our spherical model calculations for DGP gravity for a background with $\Om=1$, $\Ov=0$ and an expansion given by the standard gravity expectation (Einstein-de-Sitter). The solid lines show the calculation for $H_0 r_\mathrm{c}=0.57$ (red), $1.2$ (green) and $5.65$ (blue) while the dashed lines shows our fitting function. The solid black lines show the standard gravity expectation of $\dc\simeq1.686$ and $\Dv\simeq178$, which is independent of collapse time.}
\label{fig:DGP_spherical}
\end{figure}

In the case of DGP we include the factor of $1+1/3\beta$ in the right-hand sides of the equations, which comes from the linear DGP scalar-field equation. We do not explicitly account for the screening mechanism in our calculation. We show results for $\Om=1$ DGP models in Fig.~\ref{fig:DGP_spherical} together with fitting formulae for $\dc$ and $\Dv$ that were calibrated to the results of this calculation. The fitting function for $\Dv$ is a power law, but that for $\dc$ had to be fit via a sigmoid because of the `S' shaped bend visible in the most extreme model in Fig.~\ref{fig:DGP_spherical}. The fitting functions are
\begin{equation}
\eqalign{
%\dc&\simeq 1.686\times S(1.,0.9973,\log_{10}[0.43(H_0r_\mathrm{c})^{0.65}],0.4,\log_{10}a)\ , \cr
\dc&\simeq 1.686\times S(\log_{10}[2.33a/(H_0r_\mathrm{c})^{0.65}],0.4,1.,0.9973)\ , \cr
\Dv&\simeq 178\times[1-0.08a(H_0 r_\mathrm{c})^{-0.7}]\ ,
}
\label{eq:spherical_nDGP_appendix}
\end{equation}
where the sigmoid is defined in equation~(\ref{eq:sigmoid}).

Note that our integration methods display some numerical noise in the results for $\dc$ and $\Dv$, which manifests in the non-smooth curves seen in Fig.~\ref{fig:DGP_spherical}, particularly for $\dc$. However, we checked our $\dc$ calculations for \LCDM models against the fitting formula of \cite{Nakamura1997} and find excellent agreement, and no indication that our calculation is biased. Our results also agree with $\dc$ and $\Dv$ values given in \cite{Schmidt2010a} for the DGP model also agree with those in \cite{Schmidt2009a} for a gravity enhancement of $1/3$.

\end{document}